\PassOptionsToPackage{dvipsnames}{xcolor}

\documentclass[journal=aelccm,manuscript=article,layout=twocolumn]{achemso}

\usepackage{amsmath,amssymb,amsfonts}
\usepackage[osf]{newpxtext}
\usepackage{newpxmath}
\usepackage[utf8]{inputenx}
\usepackage{graphicx}
\usepackage[dvipsnames]{xcolor}
\usepackage[pdftex]{hyperref}
\usepackage{braket}
\usepackage{xspace}
\usepackage{soul}
\usepackage[symbol]{footmisc}
\setfnsymbol{lamport*}
\usepackage{siunitx}

\usepackage[pdftex]{hyperref}
\hypersetup{
	pdfauthor={Duc V. Dinh},
	bookmarksnumbered=true,
	pdftitle={Rock-salt ScN(113) layers grown on AlN(11\=22) by plasma-assisted molecular beam epitaxy},
	colorlinks,
	citecolor=blue,
	linkcolor=blue,
	urlcolor=blue
}

\author{Duc V. Dinh}
\email{duc.vandinh@pdi-berlin.de}
\affiliation{Paul-Drude-Institut für Festkörperelektronik, Leibniz-Institut im Forschungsverbund Berlin e.V., Hausvogteiplatz 5--7, 10117 Berlin, Germany}

\author{Esperanza Luna}
\author{Oliver Brandt}
\affiliation{Paul-Drude-Institut für Festkörperelektronik, Leibniz-Institut im Forschungsverbund Berlin e.V., Hausvogteiplatz 5--7, 10117 Berlin, Germany}

\title[Rock-salt ScN(113) layers grown on AlN(11\=22) by plasma-assisted molecular beam epitaxy]{Rock-salt ScN(113) layers grown on AlN(11\=22) by plasma-assisted molecular beam epitaxy}


\begin{document}





\begin{abstract}
\textbf{\small
Transition-metal nitrides constitute a versatile class of materials with diverse properties and wide-ranging applications. Exploring new surface orientations and uncovering novel properties can enable innovative material configurations with tailored functionalities for device integration. Here, we report the growth and characterization of (85--210)-nm-thick undoped ScN layers on AlN(11\=22)/Al$_\mathbf{2}$O$_\mathbf{3}$(10\=10) templates via plasma-assisted molecular beam epitaxy. X-ray diffractometry and transmission electron microscopy confirm a pure (113) surface orientation with rotational twins. Two distinct in-plane relationships between ScN(113) and AlN(11\=22) have been identified: the dominant [1\=10]$_\text{ScN}$ |\!\!| [\=1\=123]$_\text{AlN}$ and [33\=2]$_\text{ScN}$ |\!\!| [1\=100]$_\text{AlN}$ (under tensile-compression), and the less prevalent [\=1\=21]$_\text{ScN}$ |\!\!| [1\=100]$_\text{AlN}$ and [7\=4\=1]$_\text{ScN}$ |\!\!| [\=1\=123]$_\text{AlN}$ (under biaxial compression). Broad photoluminescence spectra with a peak emission energy of $\mathbf{\approx 2.16}$\,eV originate from the lowest direct gap at the $\mathbf{X}$ point of the ScN band structure. Temperature-dependent Hall-effect measurements (4--380\,K) reveal that impurity band conduction dominates. The electron mobility is primarily limited by optical phonon scattering, characterized by an effective phonon energy of $\mathbf{(60 \pm 3)}$\,meV.
}  
\end{abstract}
                  


\section{Introduction}

Transition-metal group-3 (or -III$_\text{B}$) nitrides (TMNs), such as ScN, YN, NbN, HfN, etc., constitute a versatile class of materials with diverse properties and applications.\cite{Eklund2016May,Abghoui2017Nov,Biswas2019Feb,Ologunagba2022May,Mahadik2022Jul,Meng2022Aug} ScN and other TMNs are refractory materials combining high mechanical strength (e.\,g., the hardness of ScN $\approx 21$\,GPa\cite{Gall1998Dec}) and high-temperature stability (e.\,g., the melting point of ScN $\approx 2600$\,°C\cite{Gschneidner1961}), making them suitable for applications in harsh environments. Theoretical studies conducted for the (001) and (111) surfaces of ScN (together with other rocksalt TMNs like HfN, MoN, and TaN) suggest that ScN, either alone\cite{Abghoui2017Nov} or combined with noble metals like Pt and Pd,\cite{Ologunagba2022May} is a promising candidate for low-cost catalysts for the hydrogen evolution reaction, contributing to renewable and environmental sustainability. ScN has also been utilized in combination with other TMNs, such as ZrN/ScN, HfN/ScN and (Zr,W)N/ScN, for Schottky diode-based thermionic energy conversion.\cite{Rawat2009Jan,Garbrecht2016Sep} Additionally, the observed high thermoelectric efficiency suggests that ScN is a promising candidate for high-temperature thermoelectric applications.\cite{Kerdsongpanya2011Dec,Burmistrova2013Apr,Rao2020Apr} ScN has been found to have a high Mn solubility, making it a good candidate for magnetic semiconductors.\cite{A.L.-Brithen2004Oct,Herwadkar2005Dec,Saha2013Aug} Furthermore, the high refractive index and strong carrier absorption of ScN at the infrared spectrum region make it well-suited for a wide range of optoelectronic applications,\cite{Maurya2022Jul,Dinh2023Sep} offering opportunities for advancements in infrared imaging, communication, sensing, and energy harvesting technologies.

Despite its rocksalt crystal structure, ScN has been effectively combined with the wurtzite group-13 (or -III$_\text{A}$) nitrides for various purposes. For instance, it serves as defect filtering in GaN,\cite{Moram2007Oct,Johnston2009Apr} and it can be used as as a high-temperature ohmic contact owing to its low electrical resistivity.\cite{Casamento2019Oct,Dinh2023Dec} Additionally, ScN combined with GaN(0001) has been theoretically proposed for electronic applications, benefiting from their small lattice mismatch ($\upDelta a/a \approx 0.1$\%) and a significant polarization discontinuity at the ScN(111)/GaN(0001) hetero-interface.\cite{Adamski2019Dec} However, ScN is commonly reported as a degenerate semiconductor due to the presence of defects and impurities, limiting this application.\cite{Dismukes1972May,Moram2008Oct,Burmistrova2013Apr,Casamento2019Oct,Ohgaki2013Sep,Deng2015Jan,Cetnar2018Nov,Rao2020Apr,Al-Atabi2020Mar,Dinh2023Sep} Very recently, we have for the first time reported on nondegenerate ScN(111)/GaN(0001) layers with high carrier mobility.\cite{Dinh2023Dec} Finally, ScN can also integrate with AlN to form wurtzite (Sc,Al)N ternary alloys for applications in surface-acoustic-wave devices,\cite{Akiyama2009Feb,Tasnadi2010Apr,Hashimoto2013Mar,Yuan_2024} field-effect transistors,\cite{Hardy2017Apr,Wang2021Aug,Dinh2023Apr} and as novel ferroelectric material.\cite{Fichtner2019,Wang2022Jul}

\begin{figure*}[t]
	\includegraphics[width=0.7\textwidth]{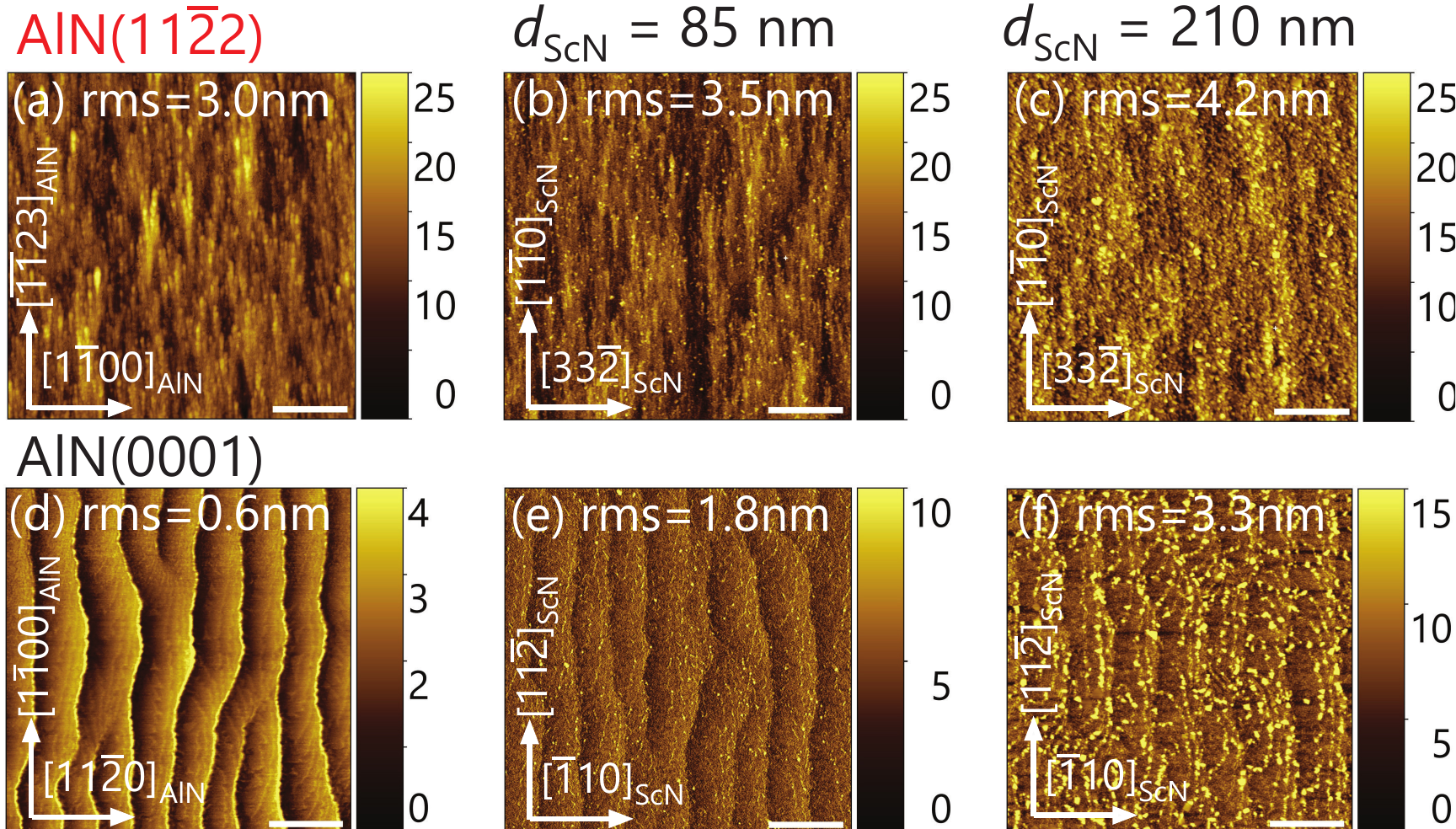}
	\caption{$5 \times 5$\,$\upmu$m$^2$ atomic force topographs of (top row) the ScN(113) layers simultaneously grown on AlN(11\=22) templates and (bottom row) the ScN(111) layers on AlN(0001) templates with different thicknesses. The scale bar is 1\,$\upmu$m, while the $z$ scale is in units of nm. Values of the root-mean-square (rms) surface roughness obtained from these topographs are given in each topograph.}
	\label{fig:AFM}
\end{figure*}

ScN layers with three different surface orientations have been previously studied, namely, ScN(001),\cite{Dismukes1972May,Gall1999Nov,Oshima2014Apr,Ohgaki2013Sep,leFebvrier2018Nov,Rao2020Apr,Moram2008May,Al-Atabi2020Mar} ScN(110),\cite{Oshima2014Apr,John2024May} and ScN(111).\cite{leFebvrier2018Nov,Casamento2019Oct,Acharya2021Nov,Dinh2023Dec,Kerdsongpanya2011Dec,Lupina2015Nov} Exploring the growth of ScN on different substrates holds great promise for uncovering new surface orientations and properties. Such investigations can lead to the development of innovative material configurations with tailored properties, thereby potentially expanding the range of applications for ScN-based devices, particularly in catalyst applications.

In this study, we investigate the structural, optical and electrical properties of undoped ScN layers simultaneously grown on AlN(11\=22)/Al$_2$O$_3$(10\=10) and AlN(0001)/Al$_2$O$_3$(0001) by plasma-assisted molecular beam epitaxy (PAMBE). X-ray diffractometry (XRD) and transmission electron microscopy (TEM) are used to determine their surface orientations and in-plane orientation-relationships. The vibrational and optical properties of the layers are investigated by Raman and photoluminescence (PL) spectroscopy, respectively. Finally, temperature-dependent Hall-effect measurements are used to elucidate their electrical properties and to shed light on the dominant scattering mechanisms.

\section{Experimental}
Intentionally undoped ScN layers are grown on semipolar AlN(11\=22) templates by PAMBE. These templates were prepared on Al$_2$O$_3$(10\=10) wafers by metal-organic vapor phase epitaxy (MOVPE).~\cite{Dinh2015Mar,Dinh2018Nov} These templates show a typical undulated surface morphology along the [1\=100]$_\text{AlN}$ direction [Fig.~\ref{fig:AFM}(a)], attributed to anisotropic diffusion of atoms along the two different in-plane directions. Before being loaded into the ultrahigh vacuum environment, the templates are cleaned in a mixture of H$_3$PO$_4$ and H$_2$SO$_4$ acids (1:3 ratio) to remove the surface oxide as well as surface contaminants, and then rinsed with de-ionized water and finally blown dry with a nitrogen gun. Prior to ScN growth, the templates were outgassed for two hours at 500\,°C in a load-lock chamber attached to the MBE system. The MBE growth chamber is equipped with high-temperature effusion cells to provide 5N-pure Sc and 6N-pure Al metals. A Veeco UNI-Bulb radio-frequency plasma source is used for the supply of active nitrogen (N$^*$). 6N-pure N$_2$ gas is used as N precursor, which is further purified by a getter filter. The N$^*$ flux is calculated from the thickness of a GaN layer grown under Ga-rich conditions, and thus with a growth rate limited by the N$^*$ flux.\cite{Dinh2023Apr,Dinh2023Dec} Prior the ScN growth, a 20-nm-thick AlN buffer layer was grown at a thermocouple temperature of 1000\,°C under slightly Al-rich conditions. Subsequently, the ScN layers are grown at the same temperature with thicknesses ($d_\text{ScN}$) ranging from 85 to 210\,nm under N$^*$-rich conditions. For comparison, ScN reference layers are grown simultaneously on polar AlN(0001) templates, which were prepared by MOVPE on Al$_2$O$_3$(0001) with a 4° miscut along the [11\=20]$_\text{AlN}$ direction.

The structural properties of the layers are characterized using a high-resolution (HR)XRD system (Philips Panalytical X'Pert PRO MRD) equipped with a two-bounce hybrid monochromator Ge(220) for the CuK$_{\alpha1}$ source ($\uplambda=\SI{1.540598}{\Angstrom}$). To further investigate the microstructural properties and details of the crystallographic arrangement at the hetero-interfaces, the 210-nm-thick ScN/AlN(11\=22) layer is selected for (scanning) transmission electron microscopy (S)TEM measurements. TEM measurements are performed with a JEOL 2100F microscope operated at 200\,kV and equipped with a bright-field (BF) detector, a high-angle annular dark-field (HAADF) detector and a Gatan Ultra Scan 4000 charge coupled device. Cross-sectional TEM specimens are prepared for observations along the orthogonal [1\=100]$_\text{AlN}$ and [\=1\=123]$_\text{AlN}$ projections of the AlN(11\=22) template using standard mechanical polishing and dimpling, followed by argon ion-milling. Analysis of high-resolution (HR)TEM phase-contrast micrographs and diffractograms is performed using Gatan Inc.'s Digital Micrograph™ (DM) software and JEMS simulation program.\cite{Stadelmann1987Jan,Stadelmann2024May}

The surface morphology of all the layers is imaged by atomic force microscopy in contact mode (Dimension Edge, Bruker). Raman and PL spectra of the layers are recorded at room temperature (RT) using a Horiba LabRAM HR Evolution\texttrademark\ Raman microscope. The samples are excited with a GaN-based diode laser $(\uplambda_{ex}=405$\,nm) and a diode-pumped solid state laser $(\uplambda_{ex}=473$\,nm) focused onto their surface for PL and Raman measurements, respectively. To investigate the electrical properties of the ScN layers, Hall-effect measurements are carried out in the van der Pauw configuration at 4--380\,K. A magnetic field of 0.7\,T is applied for these measurements.

\section{Results and Discussion}

\subsection{Surface morphology}

Figure~\ref{fig:AFM} shows the surface morphology of the ScN layers simultaneously grown on AlN(11\=22) and AlN(0001) templates. The ScN/AlN(11\=22) and ScN/AlN(0001) layers duplicate the surface morphology of the templates with slightly larger root-mean-square roughness values, that increase from 3.5 to 4.2\,nm and from 1.8 to 3.3\,nm over a scanning area of $5 \times 5$\,$\upmu$m$^2$ with increasing thickness, respectively. This increase in roughness with thickness is due to three-dimensional growth that can be clearly seen on the layers' surfaces as particle-like formations, which we attribute to the N$^*$-rich conditions.

\subsection{Structural properties}

\subsubsection{Phase purity and crystallinity}

\begin{figure}[!t]
	\includegraphics[width=0.85\columnwidth]{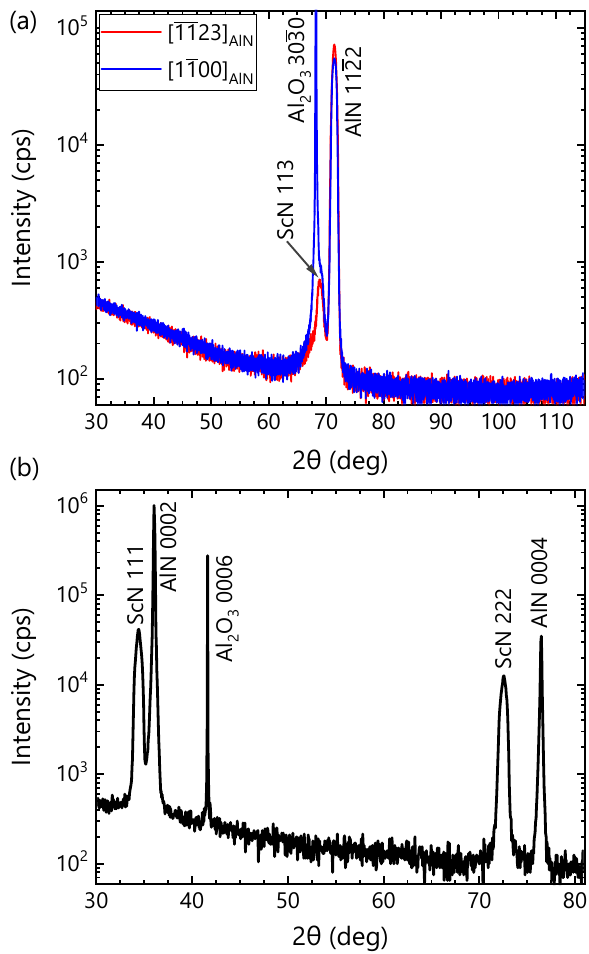}
	\caption{Symmetric 2$\uptheta-\upomega$ XRD scans of 85-nm-thick ScN layers simultaneously grown on (a) AlN(11\=22)/Al$_2$O$_3$(10\=10) and (b) AlN(0001)/Al$_2$O$_3$(0001). Scans of the ScN/AlN(11\=22) layer were performed along two in-plane directions of AlN(11\=22). All these measurements were performed with an open detector without any receiving slit.}
	\label{fig:XRD}
\end{figure}

\begin{figure}[!t]
	\includegraphics[width=0.85\columnwidth]{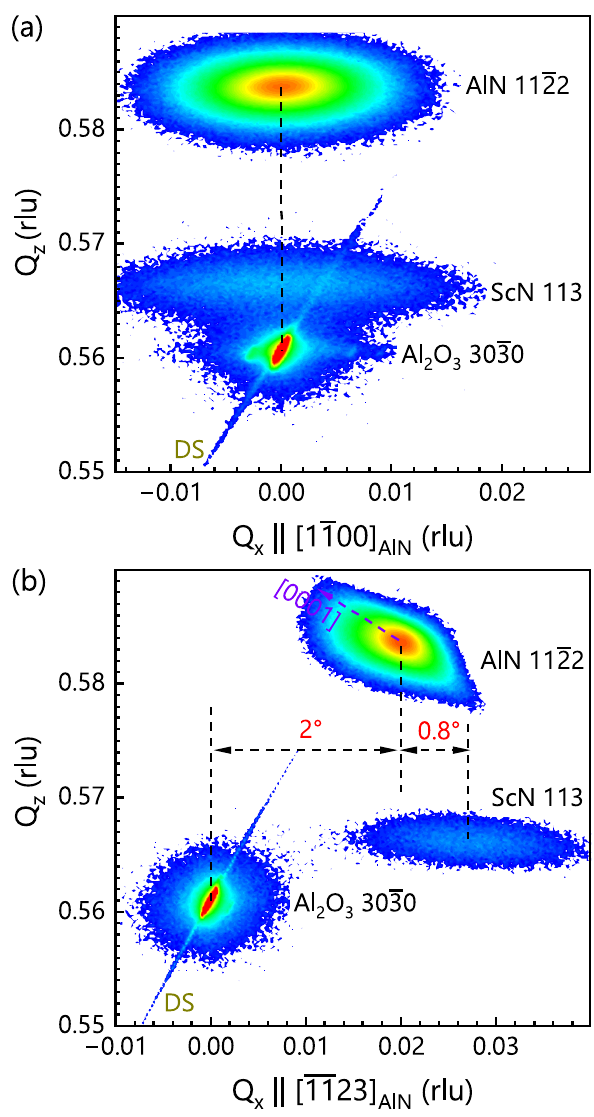}
	\caption{Reciprocal space maps of the 85-nm-thick ScN(113) layer grown on AlN(11\=22) template measured along (a) the [1\=100]$_\text{AlN}$ and (b) [\=1\=123]$_\text{AlN}$ directions. These measurements were performed with open detector using a 1-mm-wide receiving slit. In (b), the broadening along the [0001]$_\text{AlN}$ direction is due to basal-plane stacking faults.\cite{Dinh2015Mar,Dinh2018Nov} The diagonal line across the Al$_2$O$_3$ reflections are due to an instrumental artifact, namely, the detector streak (DS).}
	\label{fig:RSM}
\end{figure}

To investigate the epitaxial orientations of the layers grown on the two differently oriented AlN templates, symmetric 2$\uptheta-\upomega$ XRD survey scans has been performed over a wide angular range using an open detector. Figure~\ref{fig:XRD} shows the XRD scans of the 85-nm-thick ScN layers (data of the 210-nm-thick layers are shown in Fig.\,S1 in supplementary material). Due to the two-fold symmetry of the AlN(11\=22) template, XRD measurements of the layers grown on this template were performed along the two primary in-plane directions of AlN(11\=22). The scan along the [1\=100]$_\text{AlN}$ direction indicates two main reflections: Al$_2$O$_3$\,30\=30 and AlN\,11\=22, with an additional shoulder related to ScN observed adjacent to the Al$_2$O$_3$ reflection. When scanning along the orthogonal [\=1\=123]$_\text{AlN}$ direction, the Al$_2$O$_3$\,30\=30 reflection is not observed, while the ScN and AlN\,11\=22 reflections remain. This finding is explained by symmetric reciprocal space maps measured along these two directions, as shown in Fig.~\ref{fig:RSM}. There is almost no tilt between the ScN, AlN and Al$_2$O$_3$ reflections along the [1\=100]$_\text{AlN}$ direction, indicating that these surface planes are exactly parallel with respect to each other. A significant displacement of 2--2.8° has been observed between the reflections of AlN and ScN with respect to that of Al$_2$O$_3$ along the [\=1\=123]$_\text{AlN}$ direction. This tilt is commonly observed for nonpolar and semipolar group-III$_\text{A}$ nitrides grown on foreign substrates, and originates from the strain relaxation due to misfit dislocation at the hetero-interface.\cite{Dinh2015Mar,Wu2011Feb,DeMierry2009Mar,Dasilva2010Aug,Dinh2018Nov} The ScN 113 reflection corresponds to a lattice parameter of 4.512\,$\Angstrom$, which is identical to the value obtained for the reference layer (see Fig.\,S2 in supplementary material). Note that this lattice parameter is slightly larger than the reported value of 4.505\,$\Angstrom$ for strain-free ScN.\cite{Moram2006Jul} We attribute this increase to the high electron concentration in our layers (see their electrical properties below), leading to a lattice expansion due to the deformation potential interaction.\cite{Pietsch1983Nov,vandewalle_phys.rev.b_2003}

For the reference layers grown with different thicknesses on the AlN(0001) templates, ScN with pure (111) surface orientation is observed, as depicted in Fig.~\ref{fig:XRD}(b) for the 85-nm-thick layer [data of the 210-nm-thick layer shown in Fig.\,S1(b) in supplementary material], similar to previous reports.\cite{Casamento2019Oct,Dinh2023Sep,Dinh2023Dec,Kerdsongpanya2011Dec} The results obtained here suggest that under these identical growth conditions, both the ScN(111) and ScN(113) surfaces remain stable, with their appearances determined by the differing surface orientations of the templates. To the best of our knowledge, this is the first report ever on the growth of TMNs with (113) surface orientation.

\begin{figure}[t]
	\includegraphics[width=1\columnwidth]{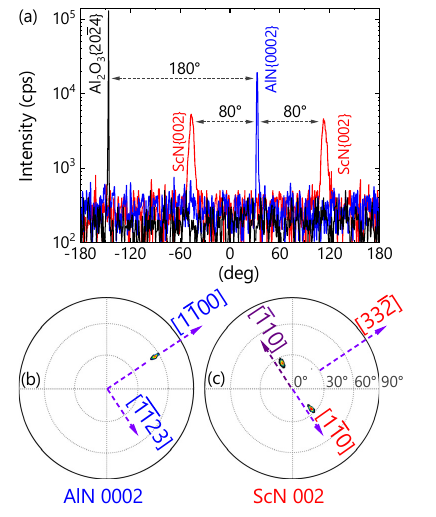}
	\caption{(a) Azimuthal scans conducted on the 85-nm-thick ScN(113) layer of the Al$_2$O$_3$\,\{20\=24\} ($\uppsi = 32.4$°), AlN \{0002\} ($\uppsi = 58.0$°) and ScN \{002\} planes ($\uppsi = 25.2$°) performed in skew-symmetric configurations. Pole figures of (b) the AlN 0002 (2$\uptheta = 36.0$°, $\upomega = 18.0$°) and (c) ScN 002 reflections (2$\uptheta = 40.0$°, $\upomega = 20.0$°).}
	\label{fig:pole113}
\end{figure}

\begin{figure}[t]
	\includegraphics[width=\columnwidth]{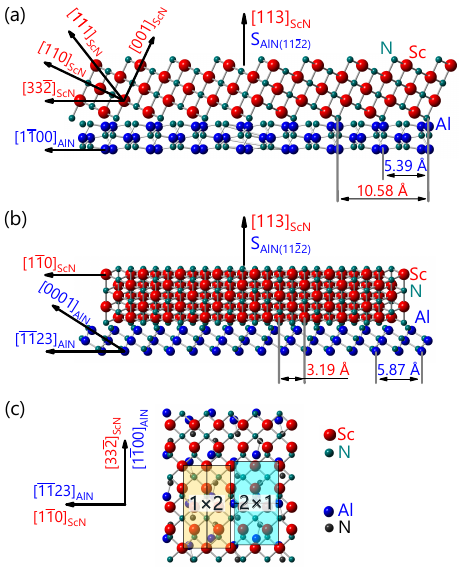}
	\caption{Side views of the atomic structures of ScN(113)/AlN(11\=22): (a) along the [33\=2]$_\text{ScN}$ and [1\=100]$_\text{AlN}$ zone axes, and (b) along the [1\=10]$_\text{ScN}$ and [\=1\=123]$_\text{AlN}$ zone axes. S$_{\mathrm{AlN}(11\bar{2}2)}$ denotes the surface normal of the AlN(11\=22) plane. (c) Top view of the ScN(113) on AlN(11\=22) atomic structures. Rectangular shapes indicate the $1 \times 2$ unit cells of ScN(113) and $2 \times 1$ unit cells of AlN(11\=22). In (c), note the different colors of N atoms on the ScN topmost and AlN underneath surfaces.}
	\label{fig:atoms}
\end{figure}

To investigate the epitaxial orientation-relationship between ScN(113), AlN(11\=22) and Al$_2$O$_3$(10\=10), azimuthal scans and pole figures have been recorded, as depicted in Fig.~\ref{fig:pole113}. These scans indicate the presence of twinning in ScN(113) with the following in-plane relationship: [1\=10]$_\text{ScN}$ |\!\!| [\=1\=123]$_\text{AlN}$ |\!\!| [0001]$_\text{Al$_2$O$_3$}$ (opposite twinned direction: [\=110]$_\text{ScN}$ |\!\!| [\=1\=123]$_\text{AlN}$) and [33\=2]$_\text{ScN}$ |\!\!| [1\=100]$_\text{AlN}$ |\!\!| [11\=20]$_\text{Al$_2$O$_3$}$, as illustrated in Fig.~\ref{fig:pole113}(c). For simplicity, we will denote the main in-plane direction of ScN as [1\=10]$_\text{ScN}$. Interestingly, a non-ideal angular spacing of $\approx$\,80°, rather than the expected 90°, between the ScN\,002 and AlN\,0002 reflections has been observed. This deviation from the ideal 90° is attributed to an in-plane lattice distortion, likely caused by residual strain and slight misorientations between twinned domains. For comparison, the ScN(111)/AlN(0001) reference layers exhibit two twin domains with the in-plane relationship [11\=2]$_\text{ScN}$ |\!\!| [1\=100]$_\text{AlN}$ and [\=110]$_\text{ScN}$ |\!\!| [1\=120]$_\text{AlN}$ (see Fig.\,S3 in supplementary material).

Figure~\ref{fig:atoms} shows ball-and-stick models of the atomic configuration of ScN(113)/AlN(11\=22) created by VESTA.\cite{VESTA} Since the surface normal of the (11\=22) plane cannot be reduced to integer Miller or Miller-Bravais indices, we denote it here by S$_{\mathrm{AlN}(11\bar{2}2)}$. Note that the angle between the [11\=22]$_\text{AlN}$ direction and S$_{\mathrm{AlN}(11\bar{2}2)}$ is 14.9°, i.\,e., the former should not be mistaken for the surface normal. 
The cross-sectional view shown in Fig.~\ref{fig:atoms}(a) indicates that the surface is stepped with \{001\} terraces and \{111\} risers, which are inclined with respect to the singular \{113\} surfaces by angles of 25.2° and 29.5°, respectively. As depicted in Fig.~\ref{fig:atoms}(c), a $1 \times 2$ super cell of ScN(113) matches a $2 \times 1$ super cell of AlN(11\=22), resulting in a near coincidence site lattice with a residual lattice mismatch ($f_\text{ScN/AlN}$) of $-1.8$\% and $8.6$\% along the [33\=2]$_\text{ScN}$ |\!\!| [1\=100]$_\text{AlN}$ direction and [1\=10]$_\text{ScN}$ |\!\!| [\=1\=123]$_\text{AlN}$ direction, respectively. Here, the $f_\text{ScN/AlN}$ values were calculated using the lattice parameter of 4.512\,$\Angstrom$ measured for the ScN layers in this study and the relaxed lattice parameters $a = 3.112$\,$\Angstrom$ and $c = 4.981$\,$\Angstrom$ reported for AlN.\cite{Paszkowicz2004Nov,Nilsson2016Apr}

\begin{figure*}[th]
	\includegraphics[width=\textwidth]{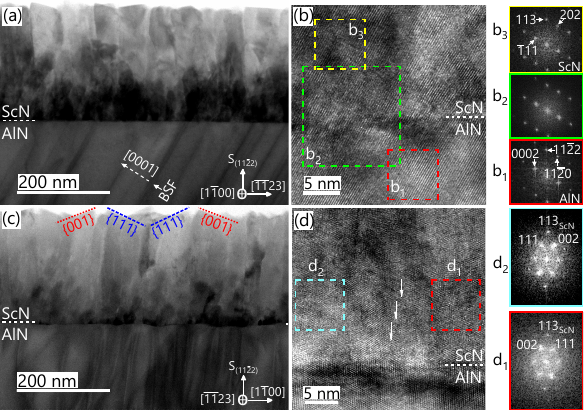}
	\caption{(a) Overview BF-STEM and (b) HRTEM images of the 210-nm-thick ScN(113) layer grown on the AlN(11\=22) template taken along the [1\=100]$_\text{AlN}$ zone axis. The inclined dark lines in the AlN area are BSFs. The square shapes in (b) indicate the areas from where the FFT patterns on (b$_1$) AlN, (b$_2$) ScN/AlN, and (b$_3$) ScN were captured. (c) Overview BF-STEM and (d) HRTEM images taken along the [\=1\=123]$_\text{AlN}$ zone axis. (d$_1$) and (d$_2$) are FFT patterns from the two selected areas in (d). Arrows in (d) indicate twin boundaries.}
	\label{fig:TEM}
\end{figure*}

X-ray rocking curve measurements were conducted to assess the crystallinity of the layers. The 85-nm-thick ScN(113) layer exhibits a slight anisotropy in ScN\,113 rocking curves measured along two in-plane directions: the values of the full-width at half maximum (FWHM) amount to 1.5° and 1.8° along the [33\=2]$_\text{ScN}$ and [1\=10]$_\text{ScN}$ directions, respectively. This is analogous to the AlN(11\=22) templates, where FWHM values of 0.3° and 0.5° are obtained in the AlN\,11\=22 rocking curves along the [\=1\=123]$_\text{AlN}$ and [1\=100]$_\text{AlN}$ directions, respectively. An anisotropic broadening of the ScN\,113 reflection is also observed in the reciprocal space maps (Fig.~\ref{fig:RSM}). In the case of the 85-nm-thick ScN(111) reference layer, the FWHM of the ScN\,111 rocking curve is 0.3°, which represents one of the best values reported in the literature.\cite{Moram2008May,Burmistrova2013Apr} The higher crystallinity of the ScN(111) layer is attributed mainly to the higher quality of the AlN(0001) templates, i.\,e., a smoother surface morphology [Fig.~\ref{fig:AFM}(d)] and higher crystallinity (FWHM values of the AlN\,0002 and AlN\,10\=13 rocking curves are 0.02° and 0.07°, respectively). Interestingly, the FWHM values of the 210-nm-thick layers are not improved compared to those of the 85-nm-thick layers. This can be explained by the insufficient lateral growth under N$^*$-rich conditions, which results in a columnar growth structure and impedes a progressive improvement in crystallinity with increasing thickness.

\subsubsection{Microstructural properties}

Overview BF-STEM images of the 210-nm-thick ScN/AlN(11\=22) layer shown in Figs.~\ref{fig:TEM}(a) and \ref{fig:TEM}(c) reveal a ScN layer composed of coalesced columnar grains with an average diameter of about 20--50\,nm. The grains have a single orientation, (113), and well-defined epitaxial relationships as will be discussed below. The basal-plane stacking faults (BSFs) in the AlN(11\=22) layer are terminated at the ScN/AlN interface [Fig.~\ref{fig:TEM}(a)], which is due to the transition from the wurtzite crystal structure of AlN to the rocksalt crystal structure of ScN. In the wurtzite structure of AlN, BSFs are defined as deviations from the ABAB... stacking sequence of the (0001) basal planes normal to the [0001]$_\text{AlN}$ direction.\cite{Dasilva2010Aug} This stacking sequence cannot propagate into the rocksalt ScN layer, which exhibits the characteristic cubic ABCABC... stacking sequence of its (111) planes normal to the [111]$_\text{ScN}$ direction.\cite{Hull2011Jan} The distinct crystallographic configurations of the wurtzite and rocksalt structures inherently inhibit the continuation of BSFs into ScN, as the rocksalt structure does not accommodate the specific planar defects found in the wurtzite structure. Consequently, this difference in crystal structure results in the natural termination of these faults at the interface between the two materials.

Figures~\ref{fig:TEM}(b) and \ref{fig:TEM}(d) display representative HRTEM phase-contrast micrographs of the ScN/AlN interface acquired along the [1\=100]$_\text{AlN}$ and [\=1\=123]$_\text{AlN}$ zone axes, respectively. Though the TEM images evidence the local presence of surface undulations at the ScN/AlN interface, there is no noticeable misalignment of the ScN(113) layer which grows epitaxially on AlN(11\=22). HRTEM investigations confirm the epitaxial relationship [1\=10]$_\text{ScN}$ |\!\!| [\=1\=123]$_\text{AlN}$ and [33\=2]$_\text{ScN}$ |\!\!| [1\=100]$_\text{AlN}$ in good agreement with XRD data. This epitaxial orientation-relationship has been determined after comparison of the experimental data with simulated diffraction patterns obtained using the JEMS simulation suite.\cite{Stadelmann1987Jan,Stadelmann2024May} Interestingly, the TEM results also reveal the presence of an additional in-plane relationship between ScN(113) and AlN(11\=22), which is [\=1\=21]$_\text{ScN}$ |\!\!| [1\=100]$_\text{AlN}$ and [7\=4\=1]$_\text{ScN}$ |\!\!| [\=1\=123]$_\text{AlN}$ (see Figs.\,S4--S5 in the supplementary material). This relationship becomes very noticeable when the layer is observed along the [1\=100]$_\text{AlN}$ zone axis (see Fig.\,S5 in the supplementary material). The additional relationship, however, has not been observed in the azimuthal and pole figure XRD measurements (Fig.~\ref{fig:pole113}). It is noteworthy that HRTEM measurements were performed on much smaller areas, typically in the nanometer range, compared to those of XRD measurements, which typically cover much larger areas (here $\approx 0.4 \times 12$\,mm$^2$). Hence, we conclude that the layer was predominantly grown with the first in-plane relationship.

Fast Fourier Transform (FFT) patterns captured on different areas on the HRTEM micrographs [Figs.~\ref{fig:TEM}(b) and \ref{fig:TEM}(d)] indicate that, regardless of the in-plane relationship, the ScN layer exhibits a (113) surface orientation, consistent with XRD data. The high density of rotational twins in the ScN(113) layer is also visible in the overall surface morphology, which is characterized by mirror-symmetric \{001\} and \{111\} facets [Figs.~\ref{fig:TEM}(a) and \ref{fig:TEM}(c)]. While both of these planes are present already on the bulk-terminated (113) surface [see Fig.~\ref{fig:atoms}(a)], their development into well-defined facets reflects the fact that they represent the crystal planes with the lowest surface energy in the rocksalt structure.\cite{Li2013Nov} 

Figure~\ref{fig:TEM}(d) shows FFT patterns captured on different areas on the HRTEM micrograph taken along the [\=1\=123]$_\text{AlN}$ zone axis and [1\=10]$_\text{ScN}$ zone axis, and following the epitaxial relationship [33\=2]$_\text{ScN}$ |\!\!| [1\=100]$_\text{AlN}$ and [1\=10]$_\text{ScN}$ |\!\!| [\=1\=123]$_\text{AlN}$. In particular, FFT patterns of the selected regions marked in Figs.~\ref{fig:TEM}(d$_1$)--\ref{fig:TEM}(d$_2$) highlight the local presence of twins, with the twinning already visible in the HRTEM micrograph itself. Additionally, the lattice parameter of the ScN layer, as determined from diffractograms generated using micrographs taken along the [1\=10]$_\text{ScN}$ |\!\!| [\=1\=123]$_\text{AlN}$ zone axis, is 4.514\,$\Angstrom$, which is very close to the value deduced using XRD.

On the other hand, diffractograms from HRTEM images along the [1\=100]$_\text{AlN}$ zone axis reveal an apparent near-perfect epitaxial alignment of the ScN and AlN lattices for grains with the epitaxial relationship [\=1\=21]$_\text{ScN}$ |\!\!| [1\=100]$_\text{AlN}$ and [7\=4\=1]$_\text{ScN}$ |\!\!| [\=1\=123]$_\text{AlN}$, as reflected in the FFT pattern in Fig.~\ref{fig:TEM}(b$2$) and in simulated diffraction patterns. Grains grown with this epitaxial relationship---[\=1\=21]$_\text{ScN}$ |\!\!| [1\=100]$_\text{AlN}$ and [7\=4\=1]$_\text{ScN}$ |\!\!| [\=1\=123]$_\text{AlN}$---are under biaxial compression along the [1\=100]$_\text{AlN}$ and [\=1\=123]$_\text{AlN}$ directions, with $f_\text{ScN/AlN} = 2.5$ and $4.0$\%, respectively (see atomic structures in Fig.\,S6 in the supplementary material).

The in-plane lattice mismatch of this secondary epitaxial alignment is lower than that of the primary epitaxial relationship, having a residual lattice mismatch of approximately $-1.8$\% and $8.6$\% along the [33\=2]$_\text{ScN}$ and [1\=10]$_\text{ScN}$ directions, respectively. At the first glance, it thus may seem that the secondary alignment should be energetically preferred. However, the elastic strain energy $U \propto C_{ij} \epsilon_i \epsilon_j$ \cite{Lifshitz1986} contains not only the normal components of the strain, but also the shear components, which we expect to be nonzero for the interface between AlN(11\=22) and ScN(113). In addition, additional factors beyond the elastic strain energy contribute to the epitaxial stabilization. These may include interfacial chemical bonding and atomic registry minimizing interface energy, anisotropic elastic properties, partial strain relaxation via dislocations, kinetic growth effects favoring certain nucleation pathways, and surface energy contributions.\cite{Zangwill1988Mar,Tersoff1994May,Zhang1997Apr} Cumulatively, these effects can stabilize an epitaxial configuration that is not at the absolute minimum in elastic strain energy.

\subsection{Vibrational and optical properties}

\begin{figure}[t]
	\includegraphics[width=\columnwidth]{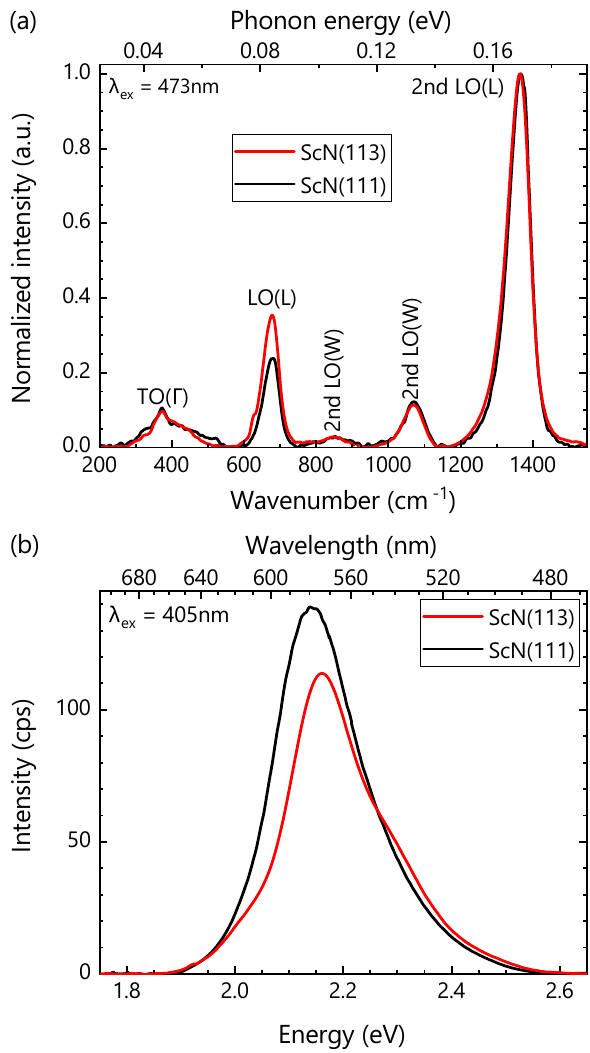}
	\caption{(a) Room-temperature Raman and (b) PL spectra of the 85-nm-thick ScN(113) and ScN(111) reference layers. } 
	\label{fig:Raman}
\end{figure}

The vibrational properties of ScN are particularly interesting due to its rocksalt structure, for which first-order Raman scattering is principally forbidden and only second-order processes are allowed. However, first-order modes are invariably observed experimentally, which is commonly attributed to a relaxation of the $k$-selection rule due to the presence of lattice disorder arsing from both point defects (such as vacancies and substitutional impurities) and structural defects.\cite{Xinh1965Nov,Todorov2011Jun} In this case, the Raman spectra basically reflects the phonon density of states (DOS), rather than consisting of only zone-center phonon modes. 

Figure~\ref{fig:Raman}(a) shows that the Raman spectra of the 85-nm-thick ScN(113) and ScN(111) layers are almost identical. Two first-order phonon modes are observed: the transverse optical (TO) mode at the $\Gamma$ point [i.e., TO($\Gamma$) at 372\,cm$^{-1}$] and the longitudinal optical (LO) mode at the $L$ point [i.e., LO(L) at $680 \pm 1$ cm$^{-1}$] in the phonon band structure of ScN. These assignments are consistent with previous results obtained on bulk ScN crystals,\cite{Travaglini1986Sep} ScN(001) layers,\cite{Maurya2022Jul} and ScN(111) layers,\cite{Lupina2015Nov,Dinh2023Sep,Dinh2023Dec} as well as with theoretical predictions.\cite{Gall2001Oct,Paudel2009Feb} As shown previously for ScN(111)/Al$_2$O$_3$(0001)\cite{Dinh2023Sep} and ScN(111)/GaN(0001)\cite{Dinh2023Dec} the spectral position of the LO($L$) mode remains unchanged with varying layer thickness, indicating that the layers investigated here are also essentially fully relaxed due to the large lattice mismatch with AlN.

In this context, the observed LO($L$) mode arises from a high phonon DOS near the $L$ point of the Brillouin zone, where the LO phonon branch is relatively flat.\cite{Paudel2009Feb,Saha2010Feb} Recently, an alternative interpretation has emerged for the rocksalt transition metal nitrides GdN and LuN.\cite{vankoughnet_phys.rev.b_2023} The authors propose that the observed LO mode is due to LO($\Gamma$) scattering induced by the Fröhlich interaction, bypassing the selection rules for deformation-potential scattering. \citet{Grumbel2024Jul} have interpreted the spectrum of bulklike ScN layers with exceptionally low electron density along the same lines. However, their Raman spectrum also exhibits the TO($X$) mode, which can only be activated by lattice disorder. Unfortunately, there is no consensus on the exact energy dispersion of the LO mode, and in particular on the ordering of the LO($\Gamma$) and LO($L$) modes, which has been shown to sensitively depend on the computational approach chosen.\cite{Li2017May} At present, it thus seems that the DOS-based interpretation provides a more consistent explanation of the spectral position and intensity of this feature in ScN, particularly for the thin layers with high defect density as considered in the present work.

Figure~\ref{fig:Raman}(b) shows an example of the PL spectra of the 85-nm-thick ScN(113) and ScN(111) layers under investigation. These layers exhibit similar PL characteristics, with a broad band centered at ($2.16 \pm 0.01$)\,eV and an FWHM of about 0.2\,eV. This peak emission aligns with the lowest direct band gap at the $X$ point of the ScN electronic band structure, consistent with previous findings.\cite{Dinh2023Sep,Lupina2015Nov,Dismukes1972May,Deng2015Jan,Moram2008Oct,Oshima2014Apr,Saha2013Aug} 

\subsection{Electrical properties}

The electrical properties of the ScN(113) and ScN(111) layers were investigated using temperature-dependent Hall-effect measurements (4--380\,K) in the van der Pauw configuration. To ensure reliable data, we focused on 85-nm-thick layers to minimize the influence of surface morphology. The ScN(113) layer exhibits a high electron density $n$ ($n_\text{RT} = 3.7 \times 10^{20}$ cm$^{-3}$) with moderate electron mobility $\mu$ ($\mu_\text{RT} = 30$ cm$^{2}$V$^{-1}$s$^{-1}$). In comparison, the ScN(111) reference layer shows a slightly lower $n$ ($n_\text{RT} = 1.1 \times 10^{20}$ cm$^{-3}$) but higher $\mu$ ($\mu_\text{RT} = 78$ cm$^{2}$V$^{-1}$s$^{-1}$). This higher $\mu$ is attributed to the higher crystallinity of the ScN(111) layer. The very high conductivity of these layers ($1-2 \times 10^5$\,S/m) highlight their potential as material for thermoelectric devices.\cite{Kerdsongpanya2011Dec,Burmistrova2013Apr,Rao2020Apr,Snyder2008Feb}

\begin{figure}[t]
	\includegraphics[width=\columnwidth]{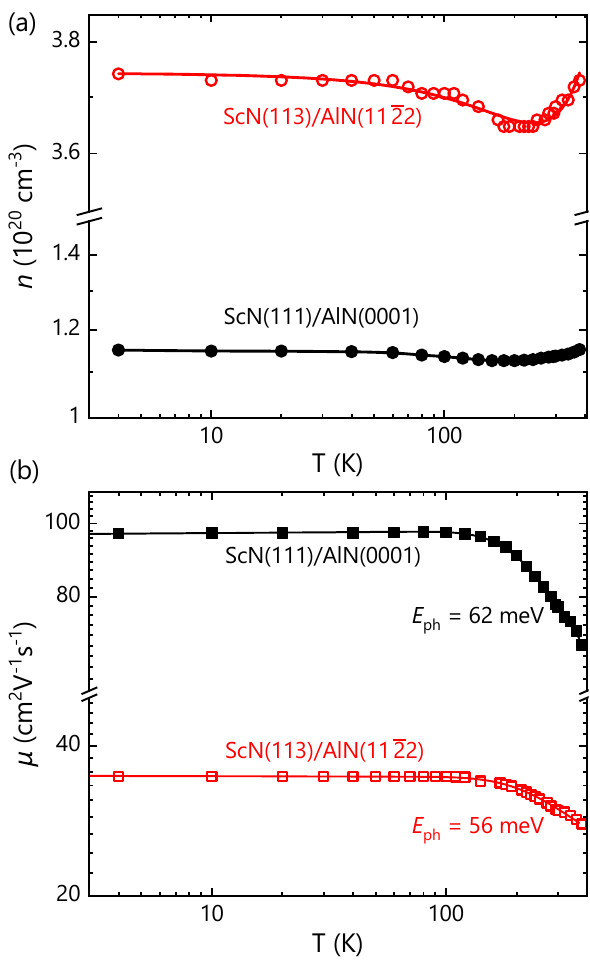}
	\caption{Temperature-dependent (a) electron concentration $n$ and (b) mobility $\mu$ of the 85-nm-thick ScN(113) and ScN(111) reference layers. Lines in (a) are guides to the eye, while lines in (b) represent the fits to the data using Eq.~\ref{eq:opt}.}
	\label{fig:Hall}
\end{figure}

For both layers, $n$ remains almost constant with temperature, with a minimum at about 250\,K, while $\mu$ is constant at low temperature and then decreases with increasing temperature. The constant values of $n$ and $\mu$ up to a certain temperature means that the layers exhibit metallic behavior, similar to previous findings reported in the literature.\cite{Dinh2023Dec,Casamento2019Oct,Cetnar2018Nov,Al-Atabi2020Mar} The entry into a semiconducting behavior at higher temperature has been observed in many semiconductors and is indicative of a gradual change from transport in an impurity- and in the conduction band at low and high temperatures, respectively.\cite{Dinh2023Dec} 

The decrease in $\mu$ with increasing temperature has been previously reported by several groups\cite{Saha2013Aug,Cetnar2018Nov,Al-Atabi2020Mar,Rao2020Apr,Dismukes1972May,Saha2017Jun} and has been qualitatively attributed to dislocation\cite{Rao2020Apr}, acoustic phonon \cite{Dismukes1972May,Saha2017Jun} or optical phonon scattering.\cite{Cetnar2018Nov,Al-Atabi2020Mar} Recently, through a quantitative analysis of the temperature-dependent $\mu$ of ScN(111) layers grown on semi-insulating GaN(0001),\cite{Dinh2023Dec} we found that at elevated temperatures, $\mu$ in these layers is primarily constrained by optical phonon scattering. This observation is consistent with recent theoretical calculations, which indicate that optical phonon scattering tends to dominate over acoustic phonon scattering in ScN.\cite{Mu2021Aug} By using the same approach, we also fit the measured $\mu(T)$ of the samples studied here using:\cite{Dinh2023Dec}
\begin{equation}
	\mu(T) = \left[A \left(e^{E_\text{ph}/k_B T}+1\right)^{-1} + B T^{-\gamma}\right]^{-1},
	\label{eq:opt}
\end{equation}
where $E_\text{ph}$ is the effective phonon energy,\cite{Askerov1975} $A$ and $B$ are constants, $k_\text{B}$ is the Boltzmann constant, and the exponent $\gamma$ represents contributions from various other scattering mechanisms including, in particular, charged point and line defects. Fit of the experimental data [Fig.~\ref{fig:Hall}(b)], return a value for $\gamma$ close to zero because of the metallic impurity-band conduction. The fits yield comparable $E_\text{ph}$ values of 56\,meV for the ScN(113) layer and 62\,meV for the ScN(111) layer. These values fall within the range of phonon energies typically observed for TO($\Gamma$) ($\approx 46$\,meV) and LO($L$) ($\approx 84$\,meV) phonon modes in ScN [see Raman spectra in Fig.~\ref{fig:Raman}(a)], consistent with our previously reported value.\cite{Dinh2023Dec} This results suggest a strong electron-optical-phonon interaction as predicted theoretically for the highly ionic semiconductor ScN.\cite{Mu2021Aug,Giustino2017Feb}

\section{Summary}

To summarize and conclude, we have reported for the first time on the growth and characterization of the ScN(113) layers grown on AlN(11\=22) templates using plasma-assisted molecular beam epitaxy. Two distinct in-plane relationships were identified: the major [1\=10]$_\text{ScN}$ |\!\!| [\=1\=123]$_\text{AlN}$ and [33\=2]$_\text{ScN}$ |\!\!| [1\=100]$_\text{AlN}$, which is under anisotropic tensile-compressive strain, and the minor [\=1\=21]$_\text{ScN}$ |\!\!| [1\=100]$_\text{AlN}$ and [7\=4\=1]$_\text{ScN}$ |\!\!| [\=1\=123]$_\text{AlN}$, which is under anisotropic biaxial compression. Room-temperature photoluminescence measurements reveal a broad luminescence band with a peak emission energy of $\approx 2.16$\,eV, corresponding to the lowest direct band gap at the $X$ point of the ScN electronic band structure. Temperature-dependent Hall-effect measurements indicate highly degenerate layers at low temperature, with transport dominated by impurity band conduction. Optical phonon scattering, with an effective phonon energy of $(60 \pm 3)$\,meV, limits carrier transport at high temperature.

\begin{acknowledgement}
We thank Carsten Stemmler for expert technical assistance with the MBE system and Doreen Steffen for TEM sample preparation. The authors are grateful to Manfred Ramsteiner for enlightening discussions and Michael Hanke for a critical reading of the manuscript. We also thank Markus Pristovsek at Nagoya University for supplying AlN templates.
\end{acknowledgement}

\begin{suppinfo}
See supplementary material for (1) symmetric 2$\uptheta-\upomega$ XRD scans of 210-nm-thick ScN layers simultaneously grown on AlN(11\=22) and AlN(0001) templates; (2) symmetric $2\uptheta-\upomega$ XRD scan and thickness simulation of the 85-nm-thick ScN(111) reference layer grown on AlN(0001) template; (3) azimuthal and pole figure scans of the 85-nm-thick ScN(111) reference layer; (4) HRTEM image of the 210-nm-thick ScN(113) grown on AlN(11\=22) template taken along the [7\=4\=1]$_\text{ScN}$ and [\=1\=123]$_\text{AlN}$ zone axes; (5) HRTEM image of the 210-nm-thick ScN(113) grown on AlN(11\=22) template taken along the [33\=2]$_\text{ScN}$ and [1\=100]$_\text{AlN}$ zone axes; (6) Side views of the atomic structures of ScN(113) on AlN(11\=22) with the in-plane relationship [\=1\=21]$_\text{ScN}$ |\!\!| [1\=100]$_\text{AlN}$ and [7\=4\=1]$_\text{ScN}$ |\!\!| [\=1\=123]$_\text{AlN}$.
\end{suppinfo}

\bibliography{2.novelScN}

\providecommand{\latin}[1]{#1}
\makeatletter
\providecommand{\doi}
  {\begingroup\let\do\@makeother\dospecials
  \catcode`\{=1 \catcode`\}=2 \doi@aux}
\providecommand{\doi@aux}[1]{\endgroup\texttt{#1}}
\makeatother
\providecommand*\mcitethebibliography{\thebibliography}
\csname @ifundefined\endcsname{endmcitethebibliography}
  {\let\endmcitethebibliography\endthebibliography}{}
\begin{mcitethebibliography}{79}
\providecommand*\natexlab[1]{#1}
\providecommand*\mciteSetBstSublistMode[1]{}
\providecommand*\mciteSetBstMaxWidthForm[2]{}
\providecommand*\mciteBstWouldAddEndPuncttrue
  {\def\EndOfBibitem{\unskip.}}
\providecommand*\mciteBstWouldAddEndPunctfalse
  {\let\EndOfBibitem\relax}
\providecommand*\mciteSetBstMidEndSepPunct[3]{}
\providecommand*\mciteSetBstSublistLabelBeginEnd[3]{}
\providecommand*\EndOfBibitem{}
\mciteSetBstSublistMode{f}
\mciteSetBstMaxWidthForm{subitem}{(\alph{mcitesubitemcount})}
\mciteSetBstSublistLabelBeginEnd
  {\mcitemaxwidthsubitemform\space}
  {\relax}
  {\relax}

\bibitem[Eklund \latin{et~al.}(2016)Eklund, Kerdsongpanya, and
  Alling]{Eklund2016May}
Eklund,~P.; Kerdsongpanya,~S.; Alling,~B. {Transition-metal-nitride-based thin
  films as novel energy harvesting materials}. \emph{J. Mater. Chem. C}
  \textbf{2016}, \emph{4}, 3905\relax
\mciteBstWouldAddEndPuncttrue
\mciteSetBstMidEndSepPunct{\mcitedefaultmidpunct}
{\mcitedefaultendpunct}{\mcitedefaultseppunct}\relax
\EndOfBibitem
\bibitem[Abghoui and Sk{\ifmmode\acute{u}\else\'{u}\fi}lason(2017)Abghoui, and
  Sk{\ifmmode\acute{u}\else\'{u}\fi}lason]{Abghoui2017Nov}
Abghoui,~Y.; Sk{\ifmmode\acute{u}\else\'{u}\fi}lason,~E. {Hydrogen evolution
  reaction catalyzed by transition-metal nitrides}. \emph{J. Phys. Chem. C}
  \textbf{2017}, \emph{121}, 24036\relax
\mciteBstWouldAddEndPuncttrue
\mciteSetBstMidEndSepPunct{\mcitedefaultmidpunct}
{\mcitedefaultendpunct}{\mcitedefaultseppunct}\relax
\EndOfBibitem
\bibitem[Biswas and Saha(2019)Biswas, and Saha]{Biswas2019Feb}
Biswas,~B.; Saha,~B. {Development of semiconducting ScN}. \emph{Phys. Rev.
  Mater.} \textbf{2019}, \emph{3}, 020301\relax
\mciteBstWouldAddEndPuncttrue
\mciteSetBstMidEndSepPunct{\mcitedefaultmidpunct}
{\mcitedefaultendpunct}{\mcitedefaultseppunct}\relax
\EndOfBibitem
\bibitem[Ologunagba and Kattel(2022)Ologunagba, and Kattel]{Ologunagba2022May}
Ologunagba,~D.; Kattel,~S. {Pt- and Pd-modified transition metal nitride
  catalysts for the hydrogen evolution reaction}. \emph{Phys. Chem. Chem.
  Phys.} \textbf{2022}, \emph{24}, 12149\relax
\mciteBstWouldAddEndPuncttrue
\mciteSetBstMidEndSepPunct{\mcitedefaultmidpunct}
{\mcitedefaultendpunct}{\mcitedefaultseppunct}\relax
\EndOfBibitem
\bibitem[Mahadik \latin{et~al.}(2022)Mahadik, Surendran, Kim, Janani, Lee, Kim,
  Kim, and Sim]{Mahadik2022Jul}
Mahadik,~S.; Surendran,~S.; Kim,~J.~Y.; Janani,~G.; Lee,~D.-K.; Kim,~T.-H.;
  Kim,~J.~K.; Sim,~U. {Syntheses and electronic structure engineering of
  transition metal nitrides for supercapacitor applications}. \emph{J. Mater.
  Chem. A} \textbf{2022}, \emph{10}, 14655\relax
\mciteBstWouldAddEndPuncttrue
\mciteSetBstMidEndSepPunct{\mcitedefaultmidpunct}
{\mcitedefaultendpunct}{\mcitedefaultseppunct}\relax
\EndOfBibitem
\bibitem[Meng \latin{et~al.}(2022)Meng, Zheng, Luo, Tang, Wang, Zhang, Tian,
  and Tang]{Meng2022Aug}
Meng,~Z.; Zheng,~S.; Luo,~R.; Tang,~H.; Wang,~R.; Zhang,~R.; Tian,~T.; Tang,~H.
  {Transition metal nitrides for electrocatalytic application: progress and
  rational design}. \emph{Nanomaterials} \textbf{2022}, \emph{12}, 2660\relax
\mciteBstWouldAddEndPuncttrue
\mciteSetBstMidEndSepPunct{\mcitedefaultmidpunct}
{\mcitedefaultendpunct}{\mcitedefaultseppunct}\relax
\EndOfBibitem
\bibitem[Gall \latin{et~al.}(1998)Gall, Petrov, Hellgren, Hultman, Sundgren,
  and Greene]{Gall1998Dec}
Gall,~D.; Petrov,~I.; Hellgren,~N.; Hultman,~L.; Sundgren,~J.~E.; Greene,~J.~E.
  {Growth of poly- and single-crystal ScN on MgO (001): Role of low-energy
  N${_2}{^+}$ irradiation in determining texture, microstructure evolution, and
  mechanical properties}. \emph{J. Appl. Phys.} \textbf{1998}, \emph{84},
  6034\relax
\mciteBstWouldAddEndPuncttrue
\mciteSetBstMidEndSepPunct{\mcitedefaultmidpunct}
{\mcitedefaultendpunct}{\mcitedefaultseppunct}\relax
\EndOfBibitem
\bibitem[Gschneidner(1961)]{Gschneidner1961}
Gschneidner,~K.~A.,~Jr. \emph{{Rare earth alloys: a critical review of the
  alloy systems of the rare earth, scandium, and yttrium metals}}; D. Van
  Nostrand Company, Ltd.: London, 1961\relax
\mciteBstWouldAddEndPuncttrue
\mciteSetBstMidEndSepPunct{\mcitedefaultmidpunct}
{\mcitedefaultendpunct}{\mcitedefaultseppunct}\relax
\EndOfBibitem
\bibitem[Rawat \latin{et~al.}(2009)Rawat, Koh, Cahill, and Sands]{Rawat2009Jan}
Rawat,~V.; Koh,~Y.~K.; Cahill,~D.~G.; Sands,~T.~D. {Thermal conductivity of
  (Zr,W)N/ScN metal/semiconductor multilayers and superlattices}. \emph{J.
  Appl. Phys.} \textbf{2009}, \emph{105}, 024909\relax
\mciteBstWouldAddEndPuncttrue
\mciteSetBstMidEndSepPunct{\mcitedefaultmidpunct}
{\mcitedefaultendpunct}{\mcitedefaultseppunct}\relax
\EndOfBibitem
\bibitem[Garbrecht \latin{et~al.}(2016)Garbrecht, Schroeder, Hultman, Birch,
  Saha, and Sands]{Garbrecht2016Sep}
Garbrecht,~M.; Schroeder,~J.~L.; Hultman,~L.; Birch,~J.; Saha,~B.; Sands,~T.~D.
  {Microstructural evolution and thermal stability of HfN/ScN, ZrN/ScN, and
  Hf$_{0.5}$Zr$_{0.5}$N/ScN metal/semiconductor superlattices}. \emph{J. Mater.
  Sci.} \textbf{2016}, \emph{51}, 8250\relax
\mciteBstWouldAddEndPuncttrue
\mciteSetBstMidEndSepPunct{\mcitedefaultmidpunct}
{\mcitedefaultendpunct}{\mcitedefaultseppunct}\relax
\EndOfBibitem
\bibitem[Kerdsongpanya \latin{et~al.}(2011)Kerdsongpanya, Van~Nong, Pryds,
  {\ifmmode\check{Z}\else\v{Z}\fi}ukauskait{\ifmmode\dot{e}\else\.{e}\fi},
  Jensen, Birch, Lu, Hultman, Wingqvist, and Eklund]{Kerdsongpanya2011Dec}
Kerdsongpanya,~S.; Van~Nong,~N.; Pryds,~N.;
  {\ifmmode\check{Z}\else\v{Z}\fi}ukauskait{\ifmmode\dot{e}\else\.{e}\fi},~A.;
  Jensen,~J.; Birch,~J.; Lu,~J.; Hultman,~L.; Wingqvist,~G.; Eklund,~P.
  {Anomalously high thermoelectric power factor in epitaxial ScN thin films}.
  \emph{Appl. Phys. Lett.} \textbf{2011}, \emph{99}, 232113\relax
\mciteBstWouldAddEndPuncttrue
\mciteSetBstMidEndSepPunct{\mcitedefaultmidpunct}
{\mcitedefaultendpunct}{\mcitedefaultseppunct}\relax
\EndOfBibitem
\bibitem[Burmistrova \latin{et~al.}(2013)Burmistrova, Maassen, Favaloro, Saha,
  Salamat, Rui~Koh, Lundstrom, Shakouri, and Sands]{Burmistrova2013Apr}
Burmistrova,~P.~V.; Maassen,~J.; Favaloro,~T.; Saha,~B.; Salamat,~S.;
  Rui~Koh,~Y.; Lundstrom,~M.~S.; Shakouri,~A.; Sands,~T.~D. {Thermoelectric
  properties of epitaxial ScN films deposited by reactive magnetron sputtering
  onto MgO(001) substrates}. \emph{J. Appl. Phys.} \textbf{2013}, \emph{113},
  153704\relax
\mciteBstWouldAddEndPuncttrue
\mciteSetBstMidEndSepPunct{\mcitedefaultmidpunct}
{\mcitedefaultendpunct}{\mcitedefaultseppunct}\relax
\EndOfBibitem
\bibitem[Rao \latin{et~al.}(2020)Rao, Biswas, Flores, Chatterjee, Garbrecht,
  Koh, Bhatia, Pillai, Hopkins, Martin-Gonzalez, and Saha]{Rao2020Apr}
Rao,~D.; Biswas,~B.; Flores,~E.; Chatterjee,~A.; Garbrecht,~M.; Koh,~Y.~R.;
  Bhatia,~V.; Pillai,~A. I.~K.; Hopkins,~P.~E.; Martin-Gonzalez,~M.; Saha,~B.
  {High mobility and high thermoelectric power factor in epitaxial ScN thin
  films deposited with plasma-assisted molecular beam epitaxy}. \emph{Appl.
  Phys. Lett.} \textbf{2020}, \emph{116}, 152103\relax
\mciteBstWouldAddEndPuncttrue
\mciteSetBstMidEndSepPunct{\mcitedefaultmidpunct}
{\mcitedefaultendpunct}{\mcitedefaultseppunct}\relax
\EndOfBibitem
\bibitem[Al-Brithen \latin{et~al.}(2004)Al-Brithen, Yang, and
  Smith]{A.L.-Brithen2004Oct}
Al-Brithen,~H.~A.; Yang,~H.; Smith,~A.~R. {Incorporation of manganese into
  semiconducting ScN using radio frequency molecular beam epitaxy}. \emph{J.
  Appl. Phys.} \textbf{2004}, \emph{96}, 3787\relax
\mciteBstWouldAddEndPuncttrue
\mciteSetBstMidEndSepPunct{\mcitedefaultmidpunct}
{\mcitedefaultendpunct}{\mcitedefaultseppunct}\relax
\EndOfBibitem
\bibitem[Herwadkar and Lambrecht(2005)Herwadkar, and
  Lambrecht]{Herwadkar2005Dec}
Herwadkar,~A.; Lambrecht,~W. R.~L. {Mn-doped $\mathrm{ScN}$: A dilute
  ferromagnetic semiconductor with local exchange coupling}. \emph{Phys. Rev.
  B} \textbf{2005}, \emph{72}, 235207\relax
\mciteBstWouldAddEndPuncttrue
\mciteSetBstMidEndSepPunct{\mcitedefaultmidpunct}
{\mcitedefaultendpunct}{\mcitedefaultseppunct}\relax
\EndOfBibitem
\bibitem[Saha \latin{et~al.}(2013)Saha, Naik, Drachev, Boltasseva, Marinero,
  and Sands]{Saha2013Aug}
Saha,~B.; Naik,~G.; Drachev,~V.~P.; Boltasseva,~A.; Marinero,~E.~E.;
  Sands,~T.~D. {Electronic and optical properties of ScN and (Sc,Mn)N thin
  films deposited by reactive DC-magnetron sputtering}. \emph{J. Appl. Phys.}
  \textbf{2013}, \emph{114}, 063519\relax
\mciteBstWouldAddEndPuncttrue
\mciteSetBstMidEndSepPunct{\mcitedefaultmidpunct}
{\mcitedefaultendpunct}{\mcitedefaultseppunct}\relax
\EndOfBibitem
\bibitem[Maurya \latin{et~al.}(2022)Maurya, Rao, Acharya, Rao, Pillai,
  Selvaraja, Garbrecht, and Saha]{Maurya2022Jul}
Maurya,~K.~C.; Rao,~D.; Acharya,~S.; Rao,~P.; Pillai,~A. I.~K.;
  Selvaraja,~S.~K.; Garbrecht,~M.; Saha,~B. {Polar semiconducting scandium
  nitride as an infrared plasmon and phonon{\textendash}polaritonic material}.
  \emph{Nano Lett.} \textbf{2022}, \emph{22}, 5182\relax
\mciteBstWouldAddEndPuncttrue
\mciteSetBstMidEndSepPunct{\mcitedefaultmidpunct}
{\mcitedefaultendpunct}{\mcitedefaultseppunct}\relax
\EndOfBibitem
\bibitem[Dinh \latin{et~al.}(2023)Dinh, Peiris,
  L{\ifmmode\ddot{a}\else\"{a}\fi}hnemann, and Brandt]{Dinh2023Sep}
Dinh,~D.~V.; Peiris,~F.; L{\ifmmode\ddot{a}\else\"{a}\fi}hnemann,~J.;
  Brandt,~O. {Optical properties of ScN layers grown on Al$_2$O$_3$(0001) by
  plasma-assisted molecular beam epitaxy}. \emph{Appl. Phys. Lett.}
  \textbf{2023}, \emph{123}, 112102\relax
\mciteBstWouldAddEndPuncttrue
\mciteSetBstMidEndSepPunct{\mcitedefaultmidpunct}
{\mcitedefaultendpunct}{\mcitedefaultseppunct}\relax
\EndOfBibitem
\bibitem[Moram \latin{et~al.}(2007)Moram, Zhang, Kappers, Barber, and
  Humphreys]{Moram2007Oct}
Moram,~M.~A.; Zhang,~Y.; Kappers,~M.~J.; Barber,~Z.~H.; Humphreys,~C.~J.
  {Dislocation reduction in gallium nitride films using scandium nitride
  interlayers}. \emph{Appl. Phys. Lett.} \textbf{2007}, \emph{91}, 152101\relax
\mciteBstWouldAddEndPuncttrue
\mciteSetBstMidEndSepPunct{\mcitedefaultmidpunct}
{\mcitedefaultendpunct}{\mcitedefaultseppunct}\relax
\EndOfBibitem
\bibitem[Johnston \latin{et~al.}(2009)Johnston, Moram, Kappers, and
  Humphreys]{Johnston2009Apr}
Johnston,~C.~F.; Moram,~M.~A.; Kappers,~M.~J.; Humphreys,~C.~J. {Defect
  reduction in (11\=22) semipolar GaN grown on m-plane sapphire using ScN
  interlayers}. \emph{Appl. Phys. Lett.} \textbf{2009}, \emph{94}, 161109\relax
\mciteBstWouldAddEndPuncttrue
\mciteSetBstMidEndSepPunct{\mcitedefaultmidpunct}
{\mcitedefaultendpunct}{\mcitedefaultseppunct}\relax
\EndOfBibitem
\bibitem[Casamento \latin{et~al.}(2019)Casamento, Wright, Chaudhuri, Xing, and
  Jena]{Casamento2019Oct}
Casamento,~J.; Wright,~J.; Chaudhuri,~R.; Xing,~H.~G.; Jena,~D. {Molecular beam
  epitaxial growth of scandium nitride on hexagonal SiC, GaN, and AlN}.
  \emph{Appl. Phys. Lett.} \textbf{2019}, \emph{115}, 172101\relax
\mciteBstWouldAddEndPuncttrue
\mciteSetBstMidEndSepPunct{\mcitedefaultmidpunct}
{\mcitedefaultendpunct}{\mcitedefaultseppunct}\relax
\EndOfBibitem
\bibitem[Dinh and Brandt(2024)Dinh, and Brandt]{Dinh2023Dec}
Dinh,~D.~V.; Brandt,~O. {Electrical properties of ScN(111) layers grown on
  semi-insulating GaN(0001) by plasma-assisted molecular beam epitaxy}.
  \emph{Phys. Rev. Appl.} \textbf{2024}, \emph{22}, 014067\relax
\mciteBstWouldAddEndPuncttrue
\mciteSetBstMidEndSepPunct{\mcitedefaultmidpunct}
{\mcitedefaultendpunct}{\mcitedefaultseppunct}\relax
\EndOfBibitem
\bibitem[Adamski \latin{et~al.}(2019)Adamski, Dreyer, and Van~de
  Walle]{Adamski2019Dec}
Adamski,~N.~L.; Dreyer,~C.~E.; Van~de Walle,~C.~G. {Giant polarization charge
  density at lattice-matched GaN/ScN interfaces}. \emph{Appl. Phys. Lett.}
  \textbf{2019}, \emph{115}, 232103\relax
\mciteBstWouldAddEndPuncttrue
\mciteSetBstMidEndSepPunct{\mcitedefaultmidpunct}
{\mcitedefaultendpunct}{\mcitedefaultseppunct}\relax
\EndOfBibitem
\bibitem[Dismukes \latin{et~al.}(1972)Dismukes, Yim, and Ban]{Dismukes1972May}
Dismukes,~J.~P.; Yim,~W.~M.; Ban,~V.~S. {Epitaxial growth and properties of
  semiconducting ScN}. \emph{J. Cryst. Growth} \textbf{1972}, \emph{13},
  365\relax
\mciteBstWouldAddEndPuncttrue
\mciteSetBstMidEndSepPunct{\mcitedefaultmidpunct}
{\mcitedefaultendpunct}{\mcitedefaultseppunct}\relax
\EndOfBibitem
\bibitem[Moram \latin{et~al.}(2008)Moram, Barber, and Humphreys]{Moram2008Oct}
Moram,~M.~A.; Barber,~Z.~H.; Humphreys,~C.~J. {The effect of oxygen
  incorporation in sputtered scandium nitride films}. \emph{Thin Solid Films}
  \textbf{2008}, \emph{516}, 8569\relax
\mciteBstWouldAddEndPuncttrue
\mciteSetBstMidEndSepPunct{\mcitedefaultmidpunct}
{\mcitedefaultendpunct}{\mcitedefaultseppunct}\relax
\EndOfBibitem
\bibitem[Ohgaki \latin{et~al.}(2013)Ohgaki, Watanabe, Adachi, Sakaguchi,
  Hishita, Ohashi, and Haneda]{Ohgaki2013Sep}
Ohgaki,~T.; Watanabe,~K.; Adachi,~Y.; Sakaguchi,~I.; Hishita,~S.; Ohashi,~N.;
  Haneda,~H. {Electrical properties of scandium nitride epitaxial films grown
  on (100) magnesium oxide substrates by molecular beam epitaxy}. \emph{J.
  Appl. Phys.} \textbf{2013}, \emph{114}, 093704\relax
\mciteBstWouldAddEndPuncttrue
\mciteSetBstMidEndSepPunct{\mcitedefaultmidpunct}
{\mcitedefaultendpunct}{\mcitedefaultseppunct}\relax
\EndOfBibitem
\bibitem[Deng \latin{et~al.}(2015)Deng, Ozsdolay, Zheng, Khare, and
  Gall]{Deng2015Jan}
Deng,~R.; Ozsdolay,~B.~D.; Zheng,~P.~Y.; Khare,~S.~V.; Gall,~D. {Optical and
  transport measurement and first-principles determination of the ScN band
  gap}. \emph{Phys. Rev. B} \textbf{2015}, \emph{91}, 045104\relax
\mciteBstWouldAddEndPuncttrue
\mciteSetBstMidEndSepPunct{\mcitedefaultmidpunct}
{\mcitedefaultendpunct}{\mcitedefaultseppunct}\relax
\EndOfBibitem
\bibitem[Cetnar \latin{et~al.}(2018)Cetnar, Reed, Badescu, Vangala, Smith, and
  Look]{Cetnar2018Nov}
Cetnar,~J.~S.; Reed,~A.~N.; Badescu,~S.~C.; Vangala,~S.; Smith,~H.~A.;
  Look,~D.~C. {Electronic transport in degenerate (100) scandium nitride thin
  films on magnesium oxide substrates}. \emph{Appl. Phys. Lett.} \textbf{2018},
  \emph{113}, 192104\relax
\mciteBstWouldAddEndPuncttrue
\mciteSetBstMidEndSepPunct{\mcitedefaultmidpunct}
{\mcitedefaultendpunct}{\mcitedefaultseppunct}\relax
\EndOfBibitem
\bibitem[Al-Atabi \latin{et~al.}(2020)Al-Atabi, Zheng, Cetnar, Look, Cahill,
  and Edgar]{Al-Atabi2020Mar}
Al-Atabi,~H.; Zheng,~Q.; Cetnar,~J.~S.; Look,~D.; Cahill,~D.~G.; Edgar,~J.~H.
  {Properties of bulk scandium nitride crystals grown by physical vapor
  transport}. \emph{Appl. Phys. Lett.} \textbf{2020}, \emph{116}, 132103\relax
\mciteBstWouldAddEndPuncttrue
\mciteSetBstMidEndSepPunct{\mcitedefaultmidpunct}
{\mcitedefaultendpunct}{\mcitedefaultseppunct}\relax
\EndOfBibitem
\bibitem[Akiyama \latin{et~al.}(2009)Akiyama, Kamohara, Kano, Teshigahara,
  Takeuchi, and Kawahara]{Akiyama2009Feb}
Akiyama,~M.; Kamohara,~T.; Kano,~K.; Teshigahara,~A.; Takeuchi,~Y.;
  Kawahara,~N. {Enhancement of piezoelectric response in scandium aluminum
  nitride alloy thin films prepared by dual reactive cosputtering}. \emph{Adv.
  Mater.} \textbf{2009}, \emph{21}, 593\relax
\mciteBstWouldAddEndPuncttrue
\mciteSetBstMidEndSepPunct{\mcitedefaultmidpunct}
{\mcitedefaultendpunct}{\mcitedefaultseppunct}\relax
\EndOfBibitem
\bibitem[Tasn{\ifmmode\acute{a}\else\'{a}\fi}di
  \latin{et~al.}(2010)Tasn{\ifmmode\acute{a}\else\'{a}\fi}di, Alling,
  H{\ifmmode\ddot{o}\else\"{o}\fi}glund, Wingqvist, Birch, Hultman, and
  Abrikosov]{Tasnadi2010Apr}
Tasn{\ifmmode\acute{a}\else\'{a}\fi}di,~F.; Alling,~B.;
  H{\ifmmode\ddot{o}\else\"{o}\fi}glund,~C.; Wingqvist,~G.; Birch,~J.;
  Hultman,~L.; Abrikosov,~I.~A. {Origin of the anomalous piezoelectric response
  in wurtzite Sc$_{x}$Al$_{1-x}$N alloys}. \emph{Phys. Rev. Lett.}
  \textbf{2010}, \emph{104}, 137601\relax
\mciteBstWouldAddEndPuncttrue
\mciteSetBstMidEndSepPunct{\mcitedefaultmidpunct}
{\mcitedefaultendpunct}{\mcitedefaultseppunct}\relax
\EndOfBibitem
\bibitem[Hashimoto \latin{et~al.}(2013)Hashimoto, Sato, Teshigahara, Nakamura,
  and Kano]{Hashimoto2013Mar}
Hashimoto,~K.-y.; Sato,~S.; Teshigahara,~A.; Nakamura,~T.; Kano,~K.
  {High-performance surface acoustic wave resonators in the 1 to 3 GHz range
  using a ScAlN/6H-SiC structure}. \emph{IEEE Trans. Ultrason. Ferroelectr.
  Freq. Control} \textbf{2013}, \emph{60}, 637\relax
\mciteBstWouldAddEndPuncttrue
\mciteSetBstMidEndSepPunct{\mcitedefaultmidpunct}
{\mcitedefaultendpunct}{\mcitedefaultseppunct}\relax
\EndOfBibitem
\bibitem[Yuan \latin{et~al.}(2024)Yuan, Dinh, Mandal, Williams, Chen, Brandt,
  and Santos]{Yuan_2024}
Yuan,~M.; Dinh,~D.~V.; Mandal,~S.; Williams,~O.~A.; Chen,~Z.; Brandt,~O.;
  Santos,~P.~V. {Generation of GHz surface acoustic waves in (Sc,Al)N thin
  films grown on free-standing polycrystalline diamond wafers by
  plasma-assisted molecular beam epitaxy}. \emph{J. Phys. D: Appl. Phys.}
  \textbf{2024}, \emph{57}, 495103\relax
\mciteBstWouldAddEndPuncttrue
\mciteSetBstMidEndSepPunct{\mcitedefaultmidpunct}
{\mcitedefaultendpunct}{\mcitedefaultseppunct}\relax
\EndOfBibitem
\bibitem[Hardy \latin{et~al.}(2017)Hardy, Downey, Nepal, Storm, Katzer, and
  Meyer]{Hardy2017Apr}
Hardy,~M.~T.; Downey,~B.~P.; Nepal,~N.; Storm,~D.~F.; Katzer,~D.~S.;
  Meyer,~D.~J. {Epitaxial ScAlN grown by molecular beam epitaxy on GaN and SiC
  substrates}. \emph{Appl. Phys. Lett.} \textbf{2017}, \emph{110}, 162104\relax
\mciteBstWouldAddEndPuncttrue
\mciteSetBstMidEndSepPunct{\mcitedefaultmidpunct}
{\mcitedefaultendpunct}{\mcitedefaultseppunct}\relax
\EndOfBibitem
\bibitem[Wang \latin{et~al.}(2021)Wang, Wang, Wang, Mohanty, Diez, Wu, Sun,
  Ahmadi, and Mi]{Wang2021Aug}
Wang,~P.; Wang,~D.; Wang,~B.; Mohanty,~S.; Diez,~S.; Wu,~Y.; Sun,~Y.;
  Ahmadi,~E.; Mi,~Z. {N-polar ScAlN and HEMTs grown by molecular beam epitaxy}.
  \emph{Appl. Phys. Lett.} \textbf{2021}, \emph{119}, 082101\relax
\mciteBstWouldAddEndPuncttrue
\mciteSetBstMidEndSepPunct{\mcitedefaultmidpunct}
{\mcitedefaultendpunct}{\mcitedefaultseppunct}\relax
\EndOfBibitem
\bibitem[Dinh \latin{et~al.}(2023)Dinh,
  L{\ifmmode\ddot{a}\else\"{a}\fi}hnemann, Geelhaar, and Brandt]{Dinh2023Apr}
Dinh,~D.~V.; L{\ifmmode\ddot{a}\else\"{a}\fi}hnemann,~J.; Geelhaar,~L.;
  Brandt,~O. {Lattice parameters of Sc$_x$Al$_{1-x}$N layers grown on GaN(0001)
  by plasma-assisted molecular beam epitaxy}. \emph{Appl. Phys. Lett.}
  \textbf{2023}, \emph{122}, 152103\relax
\mciteBstWouldAddEndPuncttrue
\mciteSetBstMidEndSepPunct{\mcitedefaultmidpunct}
{\mcitedefaultendpunct}{\mcitedefaultseppunct}\relax
\EndOfBibitem
\bibitem[Fichtner \latin{et~al.}(2019)Fichtner, Wolff, Lofink, Kienle, and
  Wagner]{Fichtner2019}
Fichtner,~S.; Wolff,~N.; Lofink,~F.; Kienle,~L.; Wagner,~B. {AlScN: A III-V
  semiconductor based ferroelectric}. \emph{J. Appl. Phys.} \textbf{2019},
  \emph{125}, 114103\relax
\mciteBstWouldAddEndPuncttrue
\mciteSetBstMidEndSepPunct{\mcitedefaultmidpunct}
{\mcitedefaultendpunct}{\mcitedefaultseppunct}\relax
\EndOfBibitem
\bibitem[Wang \latin{et~al.}(2022)Wang, Wang, Mondal, and Mi]{Wang2022Jul}
Wang,~P.; Wang,~D.; Mondal,~S.; Mi,~Z. {Ferroelectric N-polar ScAlN/GaN
  heterostructures grown by molecular beam epitaxy}. \emph{Appl. Phys. Lett.}
  \textbf{2022}, \emph{121}, 023501\relax
\mciteBstWouldAddEndPuncttrue
\mciteSetBstMidEndSepPunct{\mcitedefaultmidpunct}
{\mcitedefaultendpunct}{\mcitedefaultseppunct}\relax
\EndOfBibitem
\bibitem[Gall \latin{et~al.}(1999)Gall, Petrov, Desjardins, and
  Greene]{Gall1999Nov}
Gall,~D.; Petrov,~I.; Desjardins,~P.; Greene,~J.~E. {Microstructural evolution
  and Poisson ratio of epitaxial ScN grown on TiN(001)/MgO(001) by ultrahigh
  vacuum reactive magnetron sputter deposition}. \emph{J. Appl. Phys.}
  \textbf{1999}, \emph{86}, 5524\relax
\mciteBstWouldAddEndPuncttrue
\mciteSetBstMidEndSepPunct{\mcitedefaultmidpunct}
{\mcitedefaultendpunct}{\mcitedefaultseppunct}\relax
\EndOfBibitem
\bibitem[Oshima \latin{et~al.}(2014)Oshima,
  V{\ifmmode\acute{\imath}\else\'{\i}\fi}llora, and Shimamura]{Oshima2014Apr}
Oshima,~Y.; V{\ifmmode\acute{\imath}\else\'{\i}\fi}llora,~E.~G.; Shimamura,~K.
  {Hydride vapor phase epitaxy and characterization of high-quality ScN
  epilayers}. \emph{J. Appl. Phys.} \textbf{2014}, \emph{115}, 153508\relax
\mciteBstWouldAddEndPuncttrue
\mciteSetBstMidEndSepPunct{\mcitedefaultmidpunct}
{\mcitedefaultendpunct}{\mcitedefaultseppunct}\relax
\EndOfBibitem
\bibitem[le~Febvrier \latin{et~al.}(2018)le~Febvrier, Tureson, Stilkerich,
  Greczynski, and Eklund]{leFebvrier2018Nov}
le~Febvrier,~A.; Tureson,~N.; Stilkerich,~N.; Greczynski,~G.; Eklund,~P.
  {Effect of impurities on morphology, growth mode, and thermoelectric
  properties of (111) and (001) epitaxial-like ScN films}. \emph{J. Phys. D:
  Appl. Phys.} \textbf{2018}, \emph{52}, 035302\relax
\mciteBstWouldAddEndPuncttrue
\mciteSetBstMidEndSepPunct{\mcitedefaultmidpunct}
{\mcitedefaultendpunct}{\mcitedefaultseppunct}\relax
\EndOfBibitem
\bibitem[Moram \latin{et~al.}(2008)Moram, Novikov, Kent,
  N{\ifmmode\ddot{o}\else\"{o}\fi}renberg, Foxon, and Humphreys]{Moram2008May}
Moram,~M.~A.; Novikov,~S.~V.; Kent,~A.~J.;
  N{\ifmmode\ddot{o}\else\"{o}\fi}renberg,~C.; Foxon,~C.~T.; Humphreys,~C.~J.
  {Growth of epitaxial thin films of scandium nitride on 100-oriented silicon}.
  \emph{J. Cryst. Growth} \textbf{2008}, \emph{310}, 2746\relax
\mciteBstWouldAddEndPuncttrue
\mciteSetBstMidEndSepPunct{\mcitedefaultmidpunct}
{\mcitedefaultendpunct}{\mcitedefaultseppunct}\relax
\EndOfBibitem
\bibitem[John \latin{et~al.}(2024)John, Trampert, Dinh, Spallek,
  L{\ifmmode\ddot{a}\else\"{a}\fi}hnemann, Kaganer, Geelhaar, Brandt, and
  Auzelle]{John2024May}
John,~P.; Trampert,~A.; Dinh,~D.~V.; Spallek,~D.;
  L{\ifmmode\ddot{a}\else\"{a}\fi}hnemann,~J.; Kaganer,~V.~M.; Geelhaar,~L.;
  Brandt,~O.; Auzelle,~T. {ScN/GaN(1\=100): A new platform for the epitaxy of
  twin-free metal-semiconductor heterostructures}. \emph{Nano Lett.}
  \textbf{2024}, \emph{24}, 6233\relax
\mciteBstWouldAddEndPuncttrue
\mciteSetBstMidEndSepPunct{\mcitedefaultmidpunct}
{\mcitedefaultendpunct}{\mcitedefaultseppunct}\relax
\EndOfBibitem
\bibitem[Acharya \latin{et~al.}(2021)Acharya, Chatterjee, Bhatia, Pillai,
  Garbrecht, and Saha]{Acharya2021Nov}
Acharya,~S.; Chatterjee,~A.; Bhatia,~V.; Pillai,~A. I.~K.; Garbrecht,~M.;
  Saha,~B. {Twinned growth of ScN thin films on lattice-matched GaN
  substrates}. \emph{Mater. Res. Bull.} \textbf{2021}, \emph{143}, 111443\relax
\mciteBstWouldAddEndPuncttrue
\mciteSetBstMidEndSepPunct{\mcitedefaultmidpunct}
{\mcitedefaultendpunct}{\mcitedefaultseppunct}\relax
\EndOfBibitem
\bibitem[Lupina \latin{et~al.}(2015)Lupina, Zoellner, Niermann, Dietrich,
  Capellini, Thapa, Haeberlen, Lehmann, Storck, and Schroeder]{Lupina2015Nov}
Lupina,~L.; Zoellner,~M.~H.; Niermann,~T.; Dietrich,~B.; Capellini,~G.;
  Thapa,~S.~B.; Haeberlen,~M.; Lehmann,~M.; Storck,~P.; Schroeder,~T. {Zero
  lattice mismatch and twin-free single crystalline ScN buffer layers for GaN
  growth on silicon}. \emph{Appl. Phys. Lett.} \textbf{2015}, \emph{107},
  201907\relax
\mciteBstWouldAddEndPuncttrue
\mciteSetBstMidEndSepPunct{\mcitedefaultmidpunct}
{\mcitedefaultendpunct}{\mcitedefaultseppunct}\relax
\EndOfBibitem
\bibitem[Dinh \latin{et~al.}(2015)Dinh, Conroy, Zubialevich, Petkov, Holmes,
  and Parbrook]{Dinh2015Mar}
Dinh,~D.~V.; Conroy,~M.; Zubialevich,~V.~Z.; Petkov,~N.; Holmes,~J.~D.;
  Parbrook,~P.~J. {Single phase (11\=22) AlN grown on (10\=10) sapphire by
  metalorganic vapour phase epitaxy}. \emph{J. Cryst. Growth} \textbf{2015},
  \emph{414}, 94\relax
\mciteBstWouldAddEndPuncttrue
\mciteSetBstMidEndSepPunct{\mcitedefaultmidpunct}
{\mcitedefaultendpunct}{\mcitedefaultseppunct}\relax
\EndOfBibitem
\bibitem[Dinh \latin{et~al.}(2018)Dinh, Amano, and Pristovsek]{Dinh2018Nov}
Dinh,~D.~V.; Amano,~H.; Pristovsek,~M. {MOVPE growth and high-temperature
  annealing of (10\=10) AlN layers on (10\=10) sapphire}. \emph{J. Cryst.
  Growth} \textbf{2018}, \emph{502}, 14\relax
\mciteBstWouldAddEndPuncttrue
\mciteSetBstMidEndSepPunct{\mcitedefaultmidpunct}
{\mcitedefaultendpunct}{\mcitedefaultseppunct}\relax
\EndOfBibitem
\bibitem[Stadelmann(1987)]{Stadelmann1987Jan}
Stadelmann,~P.~A. {EMS - a software package for electron diffraction analysis
  and HREM image simulation in materials science}. \emph{Ultramicroscopy}
  \textbf{1987}, \emph{21}, 131\relax
\mciteBstWouldAddEndPuncttrue
\mciteSetBstMidEndSepPunct{\mcitedefaultmidpunct}
{\mcitedefaultendpunct}{\mcitedefaultseppunct}\relax
\EndOfBibitem
\bibitem[Stadelmann()]{Stadelmann2024May}
Stadelmann,~P. \emph{{JEMS (JEMS-SWISS, Chemin Rouge 15 CH-1805 Jongny
  Switzerland, n.d.)}}; Available at: \url{https://www.jems-swiss.ch}\relax
\mciteBstWouldAddEndPuncttrue
\mciteSetBstMidEndSepPunct{\mcitedefaultmidpunct}
{\mcitedefaultendpunct}{\mcitedefaultseppunct}\relax
\EndOfBibitem
\bibitem[Wu \latin{et~al.}(2011)Wu, Tyagi, Young, Romanov, Fujito, DenBaars,
  Nakamura, and Speck]{Wu2011Feb}
Wu,~F.; Tyagi,~A.; Young,~E.~C.; Romanov,~A.~E.; Fujito,~K.; DenBaars,~S.~P.;
  Nakamura,~S.; Speck,~J.~S. {Misfit dislocation formation at heterointerfaces
  in (Al,In)GaN heteroepitaxial layers grown on semipolar free-standing GaN
  substrates}. \emph{J. Appl. Phys.} \textbf{2011}, \emph{109}, 033505\relax
\mciteBstWouldAddEndPuncttrue
\mciteSetBstMidEndSepPunct{\mcitedefaultmidpunct}
{\mcitedefaultendpunct}{\mcitedefaultseppunct}\relax
\EndOfBibitem
\bibitem[De~Mierry \latin{et~al.}(2009)De~Mierry, Guehne, Nemoz, Chenot,
  Beraudo, and Nataf]{DeMierry2009Mar}
De~Mierry,~P.; Guehne,~T.; Nemoz,~M.; Chenot,~S.; Beraudo,~E.; Nataf,~G.
  {Comparison between polar (0001) and semipolar (11\=22) nitride
  blue{\textendash}green light-emitting diodes grown on c- and m-plane sapphire
  substrates}. \emph{Jpn. J. Appl. Phys.} \textbf{2009}, \emph{48},
  031002\relax
\mciteBstWouldAddEndPuncttrue
\mciteSetBstMidEndSepPunct{\mcitedefaultmidpunct}
{\mcitedefaultendpunct}{\mcitedefaultseppunct}\relax
\EndOfBibitem
\bibitem[Dasilva \latin{et~al.}(2010)Dasilva, Chauvat, Ruterana, Lahourcade,
  Monroy, and Nataf]{Dasilva2010Aug}
Dasilva,~Y. A.~R.; Chauvat,~M.~P.; Ruterana,~P.; Lahourcade,~L.; Monroy,~E.;
  Nataf,~G. {Defect structure in heteroepitaxial semipolar (11\=22) (Ga,Al)N}.
  \emph{J. Phys.: Condens. Matter} \textbf{2010}, \emph{22}, 355802\relax
\mciteBstWouldAddEndPuncttrue
\mciteSetBstMidEndSepPunct{\mcitedefaultmidpunct}
{\mcitedefaultendpunct}{\mcitedefaultseppunct}\relax
\EndOfBibitem
\bibitem[Moram \latin{et~al.}(2006)Moram, Barber, Humphreys, Joyce, and
  Chalker]{Moram2006Jul}
Moram,~M.~A.; Barber,~Z.~H.; Humphreys,~C.~J.; Joyce,~T.~B.; Chalker,~P.~R.
  {Young{'}s modulus, Poisson{'}s ratio, and residual stress and strain in
  (111)-oriented scandium nitride thin films on silicon}. \emph{J. Appl. Phys.}
  \textbf{2006}, \emph{100}\relax
\mciteBstWouldAddEndPuncttrue
\mciteSetBstMidEndSepPunct{\mcitedefaultmidpunct}
{\mcitedefaultendpunct}{\mcitedefaultseppunct}\relax
\EndOfBibitem
\bibitem[Pietsch and Unger(1983)Pietsch, and Unger]{Pietsch1983Nov}
Pietsch,~U.; Unger,~K. {The influence of free carriers on the equilibrium
  lattice parameter of semiconductor materials}. \emph{Phys. Status Solidi A}
  \textbf{1983}, \emph{80}, 165--172\relax
\mciteBstWouldAddEndPuncttrue
\mciteSetBstMidEndSepPunct{\mcitedefaultmidpunct}
{\mcitedefaultendpunct}{\mcitedefaultseppunct}\relax
\EndOfBibitem
\bibitem[Van De~Walle(2003)]{vandewalle_phys.rev.b_2003}
Van De~Walle,~C.~G. Effects of impurities on the lattice parameters of {{GaN}}.
  \emph{Phys. Rev. B} \textbf{2003}, \emph{68}, 165209\relax
\mciteBstWouldAddEndPuncttrue
\mciteSetBstMidEndSepPunct{\mcitedefaultmidpunct}
{\mcitedefaultendpunct}{\mcitedefaultseppunct}\relax
\EndOfBibitem
\bibitem[Momma and Izumi(2011)Momma, and Izumi]{VESTA}
Momma,~K.; Izumi,~F. {VESTA 3 for three-dimensional visualization of crystal,
  volumetric and morphology data}. \emph{J. Appl. Crystallogr.} \textbf{2011},
  \emph{44}, 1272\relax
\mciteBstWouldAddEndPuncttrue
\mciteSetBstMidEndSepPunct{\mcitedefaultmidpunct}
{\mcitedefaultendpunct}{\mcitedefaultseppunct}\relax
\EndOfBibitem
\bibitem[Paszkowicz \latin{et~al.}(2004)Paszkowicz, Podsiad{\l}o, and
  Minikayev]{Paszkowicz2004Nov}
Paszkowicz,~W.; Podsiad{\l}o,~S.; Minikayev,~R. {Rietveld-refinement study of
  aluminium and gallium nitrides}. \emph{J. Alloys Compd.} \textbf{2004},
  \emph{382}, 100\relax
\mciteBstWouldAddEndPuncttrue
\mciteSetBstMidEndSepPunct{\mcitedefaultmidpunct}
{\mcitedefaultendpunct}{\mcitedefaultseppunct}\relax
\EndOfBibitem
\bibitem[Nilsson \latin{et~al.}(2016)Nilsson,
  Janz{\ifmmode\acute{e}\else\'{e}\fi}n, and
  Kakanakova-Georgieva]{Nilsson2016Apr}
Nilsson,~D.; Janz{\ifmmode\acute{e}\else\'{e}\fi}n,~E.;
  Kakanakova-Georgieva,~A. {Lattice parameters of AlN bulk, homoepitaxial and
  heteroepitaxial material}. \emph{J. Phys. D: Appl. Phys.} \textbf{2016},
  \emph{49}, 175108\relax
\mciteBstWouldAddEndPuncttrue
\mciteSetBstMidEndSepPunct{\mcitedefaultmidpunct}
{\mcitedefaultendpunct}{\mcitedefaultseppunct}\relax
\EndOfBibitem
\bibitem[Hull and Bacon(2011)Hull, and Bacon]{Hull2011Jan}
Hull,~D.; Bacon,~D.~J. \emph{{Introduction to dislocations (fifth edition)}};
  Butterworth-Heinemann: Oxford, England, UK, 2011; pp 85--107\relax
\mciteBstWouldAddEndPuncttrue
\mciteSetBstMidEndSepPunct{\mcitedefaultmidpunct}
{\mcitedefaultendpunct}{\mcitedefaultseppunct}\relax
\EndOfBibitem
\bibitem[Li \latin{et~al.}(2013)Li, Shao, Li, McCall, Wang, and
  Zhang]{Li2013Nov}
Li,~Z.; Shao,~S.; Li,~N.; McCall,~K.; Wang,~J.; Zhang,~S.~X. {Single
  crystalline nanostructures of topological crystalline insulator SnTe with
  distinct facets and morphologies}. \emph{Nano Lett.} \textbf{2013},
  \emph{13}, 5443--5448\relax
\mciteBstWouldAddEndPuncttrue
\mciteSetBstMidEndSepPunct{\mcitedefaultmidpunct}
{\mcitedefaultendpunct}{\mcitedefaultseppunct}\relax
\EndOfBibitem
\bibitem[Lifshitz \latin{et~al.}(1986)Lifshitz, Kosevich, and
  Pitaevskii]{Lifshitz1986}
Lifshitz,~E.~M.; Kosevich,~A.~M.; Pitaevskii,~L.~P. \emph{Theory of
  elasticity}; Butterworth-Heinemann, Oxford, England, UK, 1986\relax
\mciteBstWouldAddEndPuncttrue
\mciteSetBstMidEndSepPunct{\mcitedefaultmidpunct}
{\mcitedefaultendpunct}{\mcitedefaultseppunct}\relax
\EndOfBibitem
\bibitem[Zangwill(1988)]{Zangwill1988Mar}
Zangwill,~A. \emph{{Physics at surfaces}}; Cambridge University Press:
  Cambridge, England, UK, 1988\relax
\mciteBstWouldAddEndPuncttrue
\mciteSetBstMidEndSepPunct{\mcitedefaultmidpunct}
{\mcitedefaultendpunct}{\mcitedefaultseppunct}\relax
\EndOfBibitem
\bibitem[Tersoff and LeGoues(1994)Tersoff, and LeGoues]{Tersoff1994May}
Tersoff,~J.; LeGoues,~F.~K. {Competing relaxation mechanisms in strained
  layers}. \emph{Phys. Rev. Lett.} \textbf{1994}, \emph{72}, 3570--3573\relax
\mciteBstWouldAddEndPuncttrue
\mciteSetBstMidEndSepPunct{\mcitedefaultmidpunct}
{\mcitedefaultendpunct}{\mcitedefaultseppunct}\relax
\EndOfBibitem
\bibitem[Zhang and Lagally(1997)Zhang, and Lagally]{Zhang1997Apr}
Zhang,~Z.; Lagally,~M.~G. {Atomistic processes in the early stages of thin-film
  growth}. \emph{Science} \textbf{1997}, \emph{276}, 377--383\relax
\mciteBstWouldAddEndPuncttrue
\mciteSetBstMidEndSepPunct{\mcitedefaultmidpunct}
{\mcitedefaultendpunct}{\mcitedefaultseppunct}\relax
\EndOfBibitem
\bibitem[Xinh \latin{et~al.}(1965)Xinh, Maradudin, and
  Coldwell-Horsfall]{Xinh1965Nov}
Xinh,~N.~X.; Maradudin,~A.~A.; Coldwell-Horsfall,~R.~A. {Impurity induced first
  order Raman scattering of light byalkali-halide crystals}. \emph{J. Phys.}
  \textbf{1965}, \emph{26}, 717\relax
\mciteBstWouldAddEndPuncttrue
\mciteSetBstMidEndSepPunct{\mcitedefaultmidpunct}
{\mcitedefaultendpunct}{\mcitedefaultseppunct}\relax
\EndOfBibitem
\bibitem[Todorov \latin{et~al.}(2011)Todorov, Abrashev, Ivanov, Tsutsumanova,
  Marinova, Wang, and Iliev]{Todorov2011Jun}
Todorov,~N.~D.; Abrashev,~M.~V.; Ivanov,~V.~G.; Tsutsumanova,~G.~G.;
  Marinova,~V.; Wang,~Y.-Q.; Iliev,~M.~N. {Comparative Raman study of
  isostructural YCrO$_{3}$ and YMnO$_{3}$: Effects of structural distortions
  and twinning}. \emph{Phys. Rev. B} \textbf{2011}, \emph{83}, 224303\relax
\mciteBstWouldAddEndPuncttrue
\mciteSetBstMidEndSepPunct{\mcitedefaultmidpunct}
{\mcitedefaultendpunct}{\mcitedefaultseppunct}\relax
\EndOfBibitem
\bibitem[Travaglini \latin{et~al.}(1986)Travaglini, Marabelli, Monnier, Kaldis,
  and Wachter]{Travaglini1986Sep}
Travaglini,~G.; Marabelli,~F.; Monnier,~R.; Kaldis,~E.; Wachter,~P. {Electronic
  structure of ScN}. \emph{Phys. Rev. B} \textbf{1986}, \emph{34}, 3876\relax
\mciteBstWouldAddEndPuncttrue
\mciteSetBstMidEndSepPunct{\mcitedefaultmidpunct}
{\mcitedefaultendpunct}{\mcitedefaultseppunct}\relax
\EndOfBibitem
\bibitem[Gall \latin{et~al.}(2001)Gall, Stoehr, and Greene]{Gall2001Oct}
Gall,~D.; Stoehr,~M.; Greene,~J.~E. {Vibrational modes in epitaxial
  ${\mathrm{Ti}}_{1\ensuremath{-}x}{\mathrm{Sc}}_{x}\mathrm{N}(001)$ layers: An
  ab initio calculation and Raman spectroscopy study}. \emph{Phys. Rev. B}
  \textbf{2001}, \emph{64}, 174302\relax
\mciteBstWouldAddEndPuncttrue
\mciteSetBstMidEndSepPunct{\mcitedefaultmidpunct}
{\mcitedefaultendpunct}{\mcitedefaultseppunct}\relax
\EndOfBibitem
\bibitem[Paudel and Lambrecht(2009)Paudel, and Lambrecht]{Paudel2009Feb}
Paudel,~T.~R.; Lambrecht,~W. R.~L. {Calculated phonon band structure and
  density of states and interpretation of the Raman spectrum in rocksalt ScN}.
  \emph{Phys. Rev. B} \textbf{2009}, \emph{79}, 085205\relax
\mciteBstWouldAddEndPuncttrue
\mciteSetBstMidEndSepPunct{\mcitedefaultmidpunct}
{\mcitedefaultendpunct}{\mcitedefaultseppunct}\relax
\EndOfBibitem
\bibitem[Saha \latin{et~al.}(2010)Saha, Acharya, Sands, and
  Waghmare]{Saha2010Feb}
Saha,~B.; Acharya,~J.; Sands,~T.~D.; Waghmare,~U.~V. {Electronic structure,
  phonons, and thermal properties of ScN, ZrN, and HfN: A first-principles
  study}. \emph{J. Appl. Phys.} \textbf{2010}, \emph{107}, 033715\relax
\mciteBstWouldAddEndPuncttrue
\mciteSetBstMidEndSepPunct{\mcitedefaultmidpunct}
{\mcitedefaultendpunct}{\mcitedefaultseppunct}\relax
\EndOfBibitem
\bibitem[Van~Koughnet \latin{et~al.}(2023)Van~Koughnet, Trodahl,
  {Holmes-Hewett}, and Ruck]{vankoughnet_phys.rev.b_2023}
Van~Koughnet,~K.; Trodahl,~H.~J.; {Holmes-Hewett},~W.~F.; Ruck,~B.~J.
  Defect-activated versus intrinsic {{Raman}} spectra of {{GdN}} and {{LuN}}.
  \emph{Phys. Rev. B} \textbf{2023}, \emph{108}, 064306\relax
\mciteBstWouldAddEndPuncttrue
\mciteSetBstMidEndSepPunct{\mcitedefaultmidpunct}
{\mcitedefaultendpunct}{\mcitedefaultseppunct}\relax
\EndOfBibitem
\bibitem[Gr{\ifmmode\ddot{u}\else\"{u}\fi}mbel
  \latin{et~al.}(2024)Gr{\ifmmode\ddot{u}\else\"{u}\fi}mbel, Goldhahn,
  Feneberg, Oshima, Dubroka, and Ramsteiner]{Grumbel2024Jul}
Gr{\ifmmode\ddot{u}\else\"{u}\fi}mbel,~J.; Goldhahn,~R.; Feneberg,~M.;
  Oshima,~Y.; Dubroka,~A.; Ramsteiner,~M. {Band gaps and phonons of quasi-bulk
  rocksalt ScN}. \emph{Phys. Rev. Mater.} \textbf{2024}, \emph{8},
  L071601\relax
\mciteBstWouldAddEndPuncttrue
\mciteSetBstMidEndSepPunct{\mcitedefaultmidpunct}
{\mcitedefaultendpunct}{\mcitedefaultseppunct}\relax
\EndOfBibitem
\bibitem[Li and Broido(2017)Li, and Broido]{Li2017May}
Li,~C.; Broido,~D. {Phonon thermal transport in transition-metal and rare-earth
  nitride semiconductors from first principles}. \emph{Phys. Rev. B}
  \textbf{2017}, \emph{95}, 205203\relax
\mciteBstWouldAddEndPuncttrue
\mciteSetBstMidEndSepPunct{\mcitedefaultmidpunct}
{\mcitedefaultendpunct}{\mcitedefaultseppunct}\relax
\EndOfBibitem
\bibitem[Snyder and Toberer(2008)Snyder, and Toberer]{Snyder2008Feb}
Snyder,~G.~J.; Toberer,~E.~S. {Complex thermoelectric materials}. \emph{Nat.
  Mater.} \textbf{2008}, \emph{7}, 105--114\relax
\mciteBstWouldAddEndPuncttrue
\mciteSetBstMidEndSepPunct{\mcitedefaultmidpunct}
{\mcitedefaultendpunct}{\mcitedefaultseppunct}\relax
\EndOfBibitem
\bibitem[Saha \latin{et~al.}(2017)Saha, Garbrecht, Perez-Taborda, Fawey, Koh,
  Shakouri, Martin-Gonzalez, Hultman, and Sands]{Saha2017Jun}
Saha,~B.; Garbrecht,~M.; Perez-Taborda,~J.~A.; Fawey,~M.~H.; Koh,~Y.~R.;
  Shakouri,~A.; Martin-Gonzalez,~M.; Hultman,~L.; Sands,~T.~D. {Compensation of
  native donor doping in ScN: Carrier concentration control and p-type ScN}.
  \emph{Appl. Phys. Lett.} \textbf{2017}, \emph{110}, 252104\relax
\mciteBstWouldAddEndPuncttrue
\mciteSetBstMidEndSepPunct{\mcitedefaultmidpunct}
{\mcitedefaultendpunct}{\mcitedefaultseppunct}\relax
\EndOfBibitem
\bibitem[Mu \latin{et~al.}(2021)Mu, Rowberg, Leveillee, Giustino, and Van~de
  Walle]{Mu2021Aug}
Mu,~S.; Rowberg,~A. J.~E.; Leveillee,~J.; Giustino,~F.; Van~de Walle,~C.~G.
  {First-principles study of electron transport in ScN}. \emph{Phys. Rev. B}
  \textbf{2021}, \emph{104}, 075118\relax
\mciteBstWouldAddEndPuncttrue
\mciteSetBstMidEndSepPunct{\mcitedefaultmidpunct}
{\mcitedefaultendpunct}{\mcitedefaultseppunct}\relax
\EndOfBibitem
\bibitem[Askerov(1994)]{Askerov1975}
Askerov,~B.~M. \emph{{Electron transport phenomena in semiconductors}}; World
  Scientific: Singapore, 1994; pp 78--167\relax
\mciteBstWouldAddEndPuncttrue
\mciteSetBstMidEndSepPunct{\mcitedefaultmidpunct}
{\mcitedefaultendpunct}{\mcitedefaultseppunct}\relax
\EndOfBibitem
\bibitem[Giustino(2017)]{Giustino2017Feb}
Giustino,~F. {Electron-phonon interactions from first principles}. \emph{Rev.
  Mod. Phys.} \textbf{2017}, \emph{89}, 015003\relax
\mciteBstWouldAddEndPuncttrue
\mciteSetBstMidEndSepPunct{\mcitedefaultmidpunct}
{\mcitedefaultendpunct}{\mcitedefaultseppunct}\relax
\EndOfBibitem
\end{mcitethebibliography}
\end{document}